\documentclass [10pt,letterpaper]{JHEP3}
\usepackage{amsmath}
\usepackage{epsfig}
\usepackage{epstopdf}
\usepackage{psfrag}
\usepackage{cite}       
\usepackage{bm}         
\usepackage{verbatim}   
\advance\parskip 1.0pt plus 1.0pt minus 2.0pt   
\def\href#1#2{#2}	

\def\coeff#1#2{{\textstyle {\frac {#1}{#2}}}}
\def\half{\coeff 12}

\def\G{{\cal G}}

\def\R{{\mathbb R}}

\def\tr{{\rm tr}}

\def\Z{{\mathbb Z}}

\def\Dslash{{\rlap{\raise 1pt \hbox{$\>/$}}D}}

\preprint{SLAC-PUB-13221}
\title
    {%
    \boldmath  
    Topological symmetry, 
spin liquids and CFT duals  of 
Polyakov model with  massless fermions
           }
          
\author
    {%
   Mithat \"Unsal$^1$\footnote{\email{unsal@slac.stanford.edu}}
     \\${}^1$ SLAC and Physics Department, Stanford University, Stanford, CA 94309/94305
       }%
    
\abstract
    {%
 We  prove the absence of a mass gap and confinement in the Polyakov model 
  with massless complex fermions in any representation of the gauge group.  
  A $U(1)_{*}$  topological shift symmetry protects   the masslessness of one 
   dual photon. This  symmetry  emerges  in the IR as a  consequence of the  Callias index theorem and  abelian duality. 
    For  matter in the  fundamental  representation, 
    the infrared limits of this    class of theories   
   interpolate  between   weakly    and   strongly  coupled conformal field theory (CFT)  depending on the number of flavors, and provide  an infinite class of CFTs  in $d=3$ dimensions.  The long distance physics of the model is same as certain stable spin liquids. 
   Altering the topology of the adjoint Higgs field by turning it into a compact scalar does not change  the long distance dynamics in perturbation theory, however,  non-perturbative effects 
 lead  to  a mass gap for the gauge fluctuations. 
 This  provides conceptual clarity  to many subtle issues about  compact 
 QED$_3$ discussed in the context of quantum magnets, spin liquids and  phase fluctuation models  in cuprate superconductors. 
 These constructions  also provide  new insights into  zero temperature gauge theory dynamics on 
 $\R^{2,1}$ and $\R^{2,1} \times S^1$.  
 The confined versus deconfined long distance dynamics  
 is characterized  by a discrete versus continuous   topological symmetry. 
              }%
\begin{document}

\section{Introduction}
The Polyakov model,    Yang-Mills theory with an adjoint Higgs scalar  on $\R^3$, 
is one of the cornerstones in the study of confinement in gauge theories \cite{Polyakov:1976fu}. 
Abelian duality is used  to show the emergence of  a mass gap, and to exhibit linear confinement via the proliferation of the monopoles in the  vacuum. 
Another theory which realizes confinement and a mass gap similarly, i.e, via the proliferation of the  flux (or monopoles) is  compact lattice QED$_3$.  These are two  different microscopic theories with a different set of symmetries at the cut-off scale.   However, at long distances,  they are  gapped, and  they  flow to the same theory, constituting a non-perturbative long distance duality. 

Although we do not know whether the Polyakov model is relevant in Nature,  the lattice 
QED$_3$ with fermionic fields 
appears in two dimensional spin systems, in the  spin liquid approach to high $T_c$ superconductivity and in the  phase fluctuation model of the   cuprate superconductors 
(See the reviews   \cite{carlson-2002, lee-2004} and \cite{franz-2002-66}.)
 Therefore, the issue of deconfined versus confined  long distance characteristic 
 of 2+1 dimensional lattice QED with fermionic matter  is  experimentally 
 relevant.  An important question in this context is the existence (or non-perturbative stability) of the spin liquids, the non-magnetic  Mott insulators with no broken symmetries.   
 In QED$_3$, this question translates into whether the strongly coupled fermions and gauge fluctuations  remain  massless in the long distances, when the non-perturbative effects 
 (consistent with microscopic symmetries)  are taken into account. If so, this implies 
  deconfinement   and    stability.
   In the literature, a permanent  confinement and instability was argued in  \cite{sachdev-2002-298, herbut-2003-91,herbut-2003-68}. 
       Ref. \cite{hermele-2004-70, hermele-2005-72} showed that, at least in a  large $n_f$ limit where $SU(2)$ spin symmetry is generalized to $SU(n_f)$,  there are some spin liquids 
       which are stable. 
  For  small numbers of fermionic  flavors, which is experimentally most interesting,  
  this is still an unsettled matter. 
 
In this work,   we  discuss a variety of  related gauge theories, each of which  needs to be 
distinguished  very carefully via their microscopic symmetries.  For example, consider non-compact continuum QED$_3$ minimally coupled to $2n_f$ flavors of fundamental fermions, and assume one wishes to incorporate the compactness of the gauge field. 
  We show that, common bottom-up arguments which claim to account for the compactness of the   gauge fields  are     ill-defined, due to non-uniqueness of this procedure.     In the continuum, 
 a standard way to obtain compact 
  QED$_3$ is via the gauge ``symmetry breaking" $SU(N) \rightarrow U(1)^{N-1} $ in a
parent  Yang-Mills adjoint Higgs systems.  We show that there are at least two classes of parent theories  which differ in the topological structure of their adjoint Higgs field (compact versus noncompact), yet  both  lead  to the desired gauge symmetry breaking and reduce (necessarily) to continuum  compact QED$_3$. Although indistinguishable in perturbation theory, 
the non-perturbative behavior of these theories are strikingly  opposite: In the theory with non-compact  adjoint Higgs scalar (Polyakov model with massless fermions), we demonstrate   
  \begin{eqnarray}
\underbrace{SU(N)}_{\rm  with \; noncompact \; scalar} \; \; \; \underbrace{\longrightarrow}_{\rm  Higgsing} \; \; \;  \underbrace{[U(1)]^{N-1}}_{ {\rm compact \; QED}_3}  \; \; \; \underbrace{\longrightarrow}_{\rm nonperturbative}  \qquad \underbrace{U(1)}_{ \rm CFT \; or \; free \; photon}\, ,  
\label{pattern1}
  \end{eqnarray} 
  the existence of  a massless photon  in the long distances, and the absence of confinement. 
Of course, the dramatic behavior here is the appearance of a conformal field theory (CFT) in certain cases, to be discussed below.       
In the theory with compact adjoint Higgs field, the gauge structure reduces at longer distances as 
(for moderately small number of flavors) 
 \begin{eqnarray}
\underbrace{SU(N)}_{\rm with \; compact \; scalar} \;\;\;   \underbrace{\longrightarrow}_{\rm  Higgsing}  \; \; \;  \underbrace{[U(1)]^{N-1}}_{ {\rm compact \; QED}_3}  \; \; \;    \underbrace{\longrightarrow}_{\rm nonperturbative} \qquad  \underbrace{\rm nothing}_{\rm gapped \; gauge \; bosons} \, .
\label{pattern2}
  \end{eqnarray} 
  The photon  gains  a mass, and the theory confines. 
As opposed to the common assertions  in the literature, the presence or absence of monopoles 
has nothing to do with the confining or deconfining behavior of a {\it generic} gauge theory.  (See the table.\ref{default}). We introduce  a sharp  (topological)  symmetry characterization to describe the long distance limits (deconfined versus confined, and more delicate refinements) of 
gauge theories on $\R^3$ and small $S^1 \times \R^3$.  
 \footnote{
 Naively, the (\ref{pattern1}) seems to be in accord with 
 Ref.\cite{hermele-2004-70, hermele-2005-72}, and  (\ref{pattern2}) seems to be  coinciding with the 
 results of   \cite{sachdev-2002-298, herbut-2003-91,herbut-2003-68}.  This is not quite correct.  The 
 references \cite{sachdev-2002-298, herbut-2003-91,herbut-2003-68, hermele-2004-70, hermele-2005-72} study   a spin Hamiltonian which, 
   in the $\pi$-flux state,   maps into a 
 compact   lattice QED$_3$ with fermions. The global symmetries of this lattice theory 
 is different (although related, see $\S$.\ref{sec:lat}) from the continuum discussion  above.  
 Despite these differences, we will establish precise non-perturbative long distance dualities between spin system and Polyakov model with massless fermions  in certain cases.}  
 
We first discuss the question of  confinement  in the Polyakov  model with  massless fermions, either in real and complex representations.   The answer is known for one real representation  adjoint Dirac  fermion \cite{Affleck:1982as}.  The fermion number symmetry breaks down spontaneously, and there is a gapless  Nambu-Goldstone boson (the dual photon).  The   masslessness of dual photon is protected by symmetry breaking order, i.e, Goldstone theorem, and     the adjoint  fermion acquires a  mass.  
For complex representation fermions, the infrared is  more interesting. 
There are  strongly coupled   gauge fluctuations and fermions which remain massless in the infrared. 
The answer entails   a different   mechanism to keep fermions and   a boson massless. It is referred as   quantum order (or non-symmetry breaking order)   in  condensed 
matter physics  \cite{wen-2002-65, wen-2002-66}. 
 The   appearance  of quantum order in the Polyakov model is new. In the first application,  
 the spontaneous breaking of a global symmetry generates and protects a massless boson, 
 in the latter, the unbroken symmetry implies the existence of massless boson and fermions.

 The main concept behind the deconfinement in the Polyakov  model with massless fermions is 
  a  $U(1)_{*}$  {\bf topological  symmetry}. This symmetry arises in the long distance 
and     protects only one dual photon  from acquiring a mass.  It relies on the  Jackiw-Rebbi zero modes and the index theorem of Callias
\cite{Jackiw:1975fn,Callias:1977kg} . 
Due to the index theorem, a  $U(1)_A$  symmetry of the high energy theory transmutes into a   shift  symmetry for the dual photon. 
 For complex representation fermions, the combination of the topological symmetry and other global symmetries is very powerful, and  they severely  restrict  any  perturbative or  non-perturbative relevant or marginal operators that may destabilize the masslessness of the strongly interacting photon and fermions. In particular, in   theories with $N_f \geq 4 $ fundamental fermions  are quantum critical due to the absence of 
  relevant or marginal operators which may destabilize their masslessness.
  We argue that the strong correlation physics of the fermions and gauge boson at long distance produce a   scale invariant,   conformal  field theory (CFT). 
In three dimensional non-abelian gauge theories, the  earlier examples of infrared  strongly coupled CFTs are mostly among extended supersymmetric  theories \cite{Intriligator:1996ex, Kapustin:1999ha}. The nonsupersymmetric gauge theories 
discussed in this paper provide an infinite class of infrared  CFTs
which interpolate between  weak and strong coupling as the number of flavors is varied, 
$4 \leq N_f < \infty$, with  a dimensionless coupling constant $ \sim \frac{1}{\sqrt {N_f}}$.
 The $N_f=2$ theory turns out to be non-critical, due to the presence of  a relevant, non-perturbatively   generated flux  operator with fermion zero mode insertion.

The  existence of the continuous $U(1)_{*}$  topological shift symmetry is the 
{\bf necessary and  sufficient}  condition to prove that the photon remains massless in the Polyakov model with massless complex fermions.\footnote{For a  real massless Majorana fermion in the adjoint representation, there is no $U(1)_{*}$ symmetry. Such theories on $\R^3$ do indeed confine.\cite{Affleck:1982as}.} In fact,  the fundamental distinction between the theories 
 in (\ref{pattern1}) and  (\ref{pattern2}) is that, in the  latter, the continuous topological shift symmetry for the dual photon is replaced by a discrete  one.  As opposed to continuous shift symmetry, the discrete shift symmetries cannot prohibit the appearance of a mass term for the scalar.
 Thus, the photons in the latter case should acquire mass  according to symmetry considerations.  However, there is the possibility that
   the monopole fugacity may become irrelevant at large distance in the  renormalization group sense. 
   In this case, the long distance theory will exhibit an enhanced topological symmetry relative to  the microscopic theory. 
  This implies that   the presence of  the discrete  topological symmetry is 
 necessary, but not {\bf sufficient} for  confining behavior.

Finally, equipped with the understanding of the Polyakov models, we turn to the discussion of 
spin systems. As stated earlier, the spin systems can be mapped into lattice gauge   theories 
in the slave fermion mean field theory. 
We investigate  the  relation between the  Polyakov model  and  lattice  QED$_3$, both  with massless fermions,  in the  long distance limit.  These are  theories with  distinct  microscopic symmetries. But, perhaps  the most significant distinguishing feature  of the lattice QED$_3$ and continuum Polyakov models is
   the absence of an analog of the Callias index theorem in lattice QED$_3$ as shown  by Marston  \cite{Marston:1990bj} , and the analog of a global $U(1)_A$ symmetry in the lattice model.   The first is not as severe as  it sounds despite the concerns raised in literature \cite{kim-1999-272}. In fact, 
   the latter is the main problem. We will show that, were the global $U(1)_A$ a symmetry of the spin Hamiltonian, the topological symmetry would indeed  arise 
   in the infrared despite the absence of an index theorem. If this were the case,  we could have  carried 
 a precise analogy with the Polyakov model even at small $N_f$. 
     Unfortunately, only in the sufficiently  large $N_f$ limit    can we make 
 a reliable statement   about the infrared  structure of the lattice  theory.  In particular, we are not able to  improve the discussion given in \cite{hermele-2004-70, hermele-2005-72}. 
In the Polyakov model with massless fermions, we are  able to side-step 
the renormalization group  and large $N_f$ analysis of Hermele et.al. 
\cite{hermele-2004-70}. In lattice QED$_3$, this analysis seems inevitable. Thus, there is a 
 long distance duality between the spin liquids and Polyakov models with massless fermions in the large $N_f$ limit  where both theories flow into the same interacting CFT.

\section{Gauge theories in three dimensions}
 We consider    $SU(N)$  Yang-Mills gauge theory with a noncompact adjoint Higgs scalar 
 on $\R^3$  (also known as  Georgi-Glashow model) 
   in the  presence  of massless fermions. 
 The fermions are chosen in   complex and real representations such as 
 fundamenal(F) and   adjoint(adj). We will label these theories as  P(F) and P(adj),  respectively. 
 Before  discussing them, it is useful  to review the basics of the pure 
 Polyakov model \cite{Polyakov:1976fu} and set the notation.
 
\subsection{Polyakov model} 
\label{sec:Pol}
  The action of $SU(2)$ gauge theory with an  adjoint scalar  is 
\begin{eqnarray}
S= \int_{\R^3}  \;  \frac{1}{g_3^2} &&\tr \Big[ \frac{1}{4} F_{ \mu \nu}^2  +
\half (D_{\mu} \Phi)^2 
 +    V [\Phi]           \Big]
    \label{lagrangian}
\end{eqnarray}
 $\Phi$  is a Lie algebra valued non-compact scalar Higgs field, $F_{\mu \nu}$ is the 
 non-abelian  field strength, and $\mu, \nu=1,2,3$. The classical potential  $V[\Phi]$ is chosen such that,    
at tree level, the theory  is in its Higgs regime,  $SU(2) \rightarrow U(1)$. At long distances, only the abelian components are operative.  
To all orders in  perturbation theory,  the infrared is a free (non-interacting) Maxwell theory.

The Gaussian  fixed point is destabilized  due to nonperturbative instanton (monopole)   effects.  This instability is easiest to see in a dual formulation where the gauge boson is dualized to a scalar, $F= *d \sigma$.\footnote{Our discussion mostly   relies  on  symmetries.  
Therefore,  to lessen the clutter of expressions, we set the dimensionful parameters (e.g. $g_3$) 
to one.    These parameters will be restored if  necessary.}     
  Since an instanton has a finite action, they  will proliferate due to entropic effects. 
 This generates nonperturbative $e^{-S_0}$ effects in the long distance  Lagrangian  
\begin{equation}
L = \half (\partial \sigma)^2 -  e^{-S_0} (e^{i \sigma} + e^{-i \sigma})
\label{dual}
\end{equation} 
The $\cos \sigma$ is   a relevant operator    which alters the IR physics drastically,  and leads to  a  mass gap $\sim e^{-S_0/2}$. 
 
 It is worth nothing that, the dual of the free Maxwell theory, i.e., in the absence of monopoles, described by 
 $
L = \half (\partial \sigma)^2,  
$
has a continuous shift symmetry 
\begin{equation}
U(1)_{\rm flux}: \; \sigma\rightarrow \sigma - \beta
\label{U1J}
\end{equation}
which protects $\sigma$ from acquiring mass.  
  The  current associated with the shift  symmetry is 
$ {\cal J}_{\mu}= \partial_{\mu} \sigma= \half \epsilon_{\mu \nu \rho} F_{\nu \rho} = F_{\mu} $, and 
its divergence is zero, 
$  \partial_{\mu} {\cal J}_{\mu}= 
\partial_{\mu} F_{\mu}=0 $,   reflecting the absence of monopoles and conservation of magnetic flux, hence the name $U(1)_{\rm flux}$.

In the $U(1)$  gauge theory with monopoles, 
the current  ${\cal J}_{\mu} $ is not conserved. Its divergence is 
$ \partial_{\mu}  {\cal J}_{\mu}=  \nabla^2 \sigma = \partial_{\mu} F_{\mu}= \rho_m(x) $ 
where  $\rho_m(x) $ is the monopole charge density. 
 Since  the $U(1)_{\rm flux}$ is no longer a symmetry,  there  is no symmetry reason for the $\sigma$ field to remain massless. Indeed, $\sigma$ acquires a mass   as shown in 
 (\ref{dual}).
 
{$\bf SU(N)$:} More generally, let the $SU(N)$ gauge symmetry be broken down to $U(1)^{N-1}$ via an adjoint Higgs vacuum expectation value 
 \begin{eqnarray}
 \langle \Phi \rangle =  {\rm Diag}( a_1, \ldots, a_N)
 \end{eqnarray}
  where  
$  a_1 < a_2 < \ldots< a_N$.
 There are $N-1$ photons which remain massless to all orders in perturbation theory.
  Let us  dualize them into  
 $(F_1, \ldots , F_{N-1})= * d ( \sigma_1, \ldots, \sigma_{N-1})$. Non-perturbatively,  
 there are $N-1$ types of elementary monopoles  associated with this pattern, which we 
  label  by their magnetic  charges $\{{\bm \alpha}_1, \ldots, {\bm \alpha}_{N-1}\}$ where each ${\bm \alpha}_i$ is an  $N-1$  vector with charges under $U(1)^{N-1}$. The 
 antimonopoles carry  opposite charges.
  The  monopole operator in a theory without fermions  is 
  $e^{-S_0} e^{i {\bm \alpha}_i {\bm \sigma}}$, and the sum over all elementary monopole  
  effects induce  $ e^{-S_{0}}  \sum_{j=1}^{N-1}   \cos ( {\bm \alpha}_j {\bm \sigma}) $ rendering all 
 $N-1$ varieties of photons massive.  \footnote{We assume, for simplicity,  $S_{0,i} \equiv \frac{4 \pi}{g_3^2}|a_{i+1} -a_{i}| = S_0$  for the elementary  monopoles by tuning the potential. 
  This can be relaxed if desired.}

\subsubsection{Introducing  complex representation fermions} 
\label{introcomp}
Our goal is to  construct the   non-perturbative long distance description  of Polyakov models with massless fermions. The long distance effective field theory  must respect  all the (non-anomalous)  symmetries of the underlying microscopic theory.  In other words, the (perturbative or non-perturbative) operators that can be generated are 
severely restricted by the microscopic symmetries. 
 Therefore, it is useful to  clearly  state the symmetries of the microscopic P(F) model.   This will 
 also ease the comparison of microscopic and enhanced (emergent) macroscopic 
  global flavor and  spacetime  symmetries  of the theory. 
  
 Consider the addition of the  massless fermions in the fundamental representation of the gauge group into the Polyakov model. (The generalization to other complex representation fermions is possible.)  
We interchangeably use the four-component Dirac spinors or  two   two-component Dirac spinors  $\psi_1$  and $\psi_2$    related to each other via 
\begin{equation}
\Psi^a= \left(\begin{array}{l}
    \psi_1^a \\
     \bar \psi_2^a 
     \end{array}
     \right), \qquad \overline \Psi_a=    \left(\begin{array}{l}
    \psi_2^a \\
     \bar \psi_1^a 
     \end{array}
     \right),    
     \label{split}
  \end{equation}   
 We consider the theories  with $N_f=2n_f$ two component Dirac spinors, or equivalently,  
 $n_f$ four  component spinors. The  $a=1, \ldots n_f$  and subscripts $(1,2)$  are flavor indices. 
 In our conventions, the representations of the two component fermions under the $SU(N)$ gauge group are  
 $(\psi_1^a,  \bar  \psi_2^a) \in (\Box,  \Box)$ 
 where $\Box$ denotes the fundamental representation. 
  These combinations and our subsequent Dirac $\gamma$ matrix choices are 
  for later convenience, and   will make the Callias index analysis slightly simpler.  
  \footnote{In Euclidean space, $\psi_a$ and  $\bar \psi_a$ should be viewed as 
independent variables. In particular, they are not related to each other by conjugation 
}  
  The fermions couple to gauge fields and adjoint scalars as 
 \begin{equation}
   L_{\rm F}=  i \overline \Psi^a  \Big( \gamma_{\mu} (\partial_{\mu}  + i A_{\mu} )  + i \gamma_4  \Phi \Big) \Psi_a
   \end{equation} 
 where the  
 Euclidean $\gamma$ matrices are  given by 
 \begin{equation}
 \gamma_{\mu} = \sigma_1 \otimes \sigma_\mu, \; \;  \gamma_4= \sigma_2 \otimes I \qquad 
  \{\gamma_{M}, \gamma_{N}\}= 2 \delta_{MN}, \qquad M, N=1,\ldots 4
 \end{equation}
 It is also convenient to define 
 $$ \overline \sigma_M = (\sigma_{\mu}, -iI) \equiv (\sigma_{\mu}, \sigma_4), \qquad  
                          \sigma_M = (\sigma_{\mu}, iI)\equiv (\sigma_{\mu}, -\sigma_4),$$
where $\sigma_{\mu}$ are the Pauli matrices. The explicit form of the Dirac-like  operator in this basis is
\begin{eqnarray}
\gamma_M  D_M =            \gamma_{\mu}          D_{\mu} +
   \gamma_4 (i \Phi)  = \left[ \begin{array}{cc} 
0& \sigma_\mu(\partial_\mu +i A_{\mu}) + \sigma_4 (i \Phi) \cr
 \sigma_\mu(\partial_\mu +i A_{\mu}) - \sigma_4 (i \Phi) & 0
 \end{array} \right]
\end{eqnarray}
and consequently, 
 \begin{eqnarray}
   L_{\rm F}= && i \bar \psi_1^a (  \sigma_\mu(\partial_\mu +i A_{\mu}) + i \sigma_4 \Phi )\psi_1^a  \;  + \;  i  \psi_2^a (  \sigma_\mu(\partial_\mu +i A_{\mu}) - i \sigma_4  \Phi ) \bar \psi_2^a 
\label{fl}
   \end{eqnarray} 
   In this representation, it is easier to see the global symmetries of the theory.
  Besides the  $SO(3)_L $ Euclidean Lorentz symmetry and the $C, P, T$   discrete charge conjugation, parity and (Euclidean) time reversal symmetries, 
the theory possesses a discrete $\Z_2$
\begin{equation}
\Z_2: \qquad \Phi \rightarrow - \Phi,  \qquad \psi_1 \rightarrow \bar \psi_2,  \qquad \psi_2 \rightarrow \bar \psi_1  
\end{equation}
and  the following global   (flavor) symmetries
\begin{eqnarray}
\begin{array}{lll}
SU(n_f)_1: \qquad &\psi_1 \rightarrow U \psi_1, \;\;\qquad & \bar \psi_2 \rightarrow \bar \psi_2 ,  \cr \cr 
 SU(n_f)_2:  \qquad &  \psi_1 \rightarrow \psi_1, \;\; &\bar \psi_2 \rightarrow V  \bar \psi_2, 
 \cr \cr  U(1)_V: \;\;\;  \qquad & \psi_1 \rightarrow e^{i \delta}  \psi_1, \;\; & \bar \psi_2 \rightarrow  e^{ i \delta}  \bar \psi_2\cr \cr
 U(1)_A: \;\;\; \;\; \qquad & \psi_1 \rightarrow e^{i \beta}  \psi_1,  \;\;& \bar \psi_2 \rightarrow  e^{- i \beta}  \bar \psi_2
 \end{array}
\label{symfun}
\end{eqnarray}
Note that the gauge covariant term  possesses a  larger global  $SU(2n_f)$   symmetry group.  
Were the Yukawa's not present in the theory, the 
$SU(n_f)_1 \times SU(n_f)_2 \times U(1)_A$
global  symmetry would enhance into the   $SU(2n_f)$.  
However, the relative sign difference between the covariant derivative and Yukawa couplings prevents this enhancement in the microscopic theory. 
Since there is no chiral anomaly in $d=3$ dimensions,  the $U(1)_A$  symmetry 
is a  true  symmetries of the theory.  The discrete $P$ and $\Z_2$ symmetries, and continuous flavor symmetry prohibits a fermion mass term. 
 To summarize, the full  microscopic symmetry  
 group  ${\cal G_M}_{,\rm P(F)}$    of the theory is  
  \begin{eqnarray}
{\cal G_M}_{,\rm P(F)} = SO(3)_L  \times C \times P \times T  \times \Z_2 \times U(1)_V \times   U(1)_A \times 
SU(n_f)_1 \times SU(n_f)_2
\label{master1} 
\end{eqnarray}

 \subsubsection{Real representation fermions} We  restrict attention  to the  adjoint representation fermion. Since the adjoint representation is  real,  the two component (complex) Dirac spinors is appropriate for all circumstances. Thus, $N_f=n_f$. 
The coupling of fermions to gauge boson and adjoint scalar is 
 \begin{eqnarray}
   L_{\rm adj}= && i \tr \; \left[ \bar \psi_a \Big(  \sigma_\mu(\partial_\mu +i [A_{\mu}, \; ] ) + \sigma_4 [i \Phi, \; ] 
   \Big) \psi_a  \right]
   \label{f2}
   \end{eqnarray} 
The global flavor symmetries of the theory  is given by 
\begin{eqnarray}
&& SU(n_f):\;\;\; \psi \rightarrow U \psi, \cr
&& U(1)_A: \;\;\; \;\; \psi \rightarrow e^{i \beta}  \psi. 
\label{symadj}
\end{eqnarray}
Note that, in this case, $U(1)_A$ may be viewed as fermion number symmetry. However, since 
it does not have the same interpretation in the theories with complex representation fermions, 
we will not use this nomenclature. 
Thus, the full symmetry  
 group  ${\cal G_M}_{,\rm P(adj)}$    of the  microscopic   theory is  
  \begin{eqnarray}
{\cal G_M}_{,\rm P(adj)} = SO(3)_L  \times C \times P \times T \times   U(1)_A \times 
SU(n_f)
\label{master2} 
\end{eqnarray}

{\bf Remark on QCD:} At the classical level,  the flavor symmetry group of the Polyakov models with fermions on $\R^3$
  is the   same as the  flavor symmetries  of the corresponding QCD on 
  $\R^4$ or   $S^1 \times \R^3$.  However, in QCD in four dimensions, the analog of the symmetry that we referred as  $U(1)_A$  in (\ref{symfun}) and (\ref{symadj}) is  
  anomalous.  In odd dimensions, there is no chiral anomaly, and the   $U(1)_A$ is a true symmetry of  the Polyakov model with massless fermions. In four dimensions,   
  due to   instanton effects,  only a discrete 
 $\Z_{2h}$ subgroup of $U(1)_A$  survives quantization, where $2h$ is  the number of fermionic zero modes in the background of a four dimensional  instanton.   
 The microscopic   $U(1)_A$  symmetry will play a 
  major role in the characterization of deconfinement in 
 P(${\cal R}$) theories.

\subsubsection{Perturbative operators and flux operators}
\label{setup}
In all  the P$({\cal R})$ theories, we assume that the theory is always maximally Higgsed, and the long distance is dictated by the   maximal  abelian subgroup. There are massless bosons whose masslessness is protected to all orders in perturbation theory. Also, there are  fermionic zero modes which interact  with gauge fluctuations at long distances.   Our interest is to determine the stability of such massless fields.  There are two categories  of operators 
which may be generated, and alter the  long distance  physics. These are, following  
\cite{hermele-2004-70},  
\begin{itemize} 
{\item perturbative (without flux),  naturally incorporated in terms of the original  variables.}
{\item  nonperturbative (flux operators), or topological excitations, naturally incorporated in terms of dual photon.} 
\end{itemize}

For example, in the  pure Polyakov model, a would-be operator of the first category is the relevant  Chern-Simons term, 
\begin{equation}
\frac{ i n}{4 \pi} \int \epsilon^{
\mu \nu \rho}  A_{\mu} \partial_{\nu} A_\rho 
\end{equation}
which would induce a mass term for the photon. However, this operator does not get 
 generated at one loop order  (or any order in perturbation theory), because 
  the microscopic  theory is parity invariant and the   Chern-Simons term is parity odd. Thus, this type of instability does not occur. 
 
 An operator in the second category is the monopole operator. Indeed, it is allowed by all symmetries and generates the $e^{-S_0} (e^{i \sigma} + \rm c.c.)$ interaction, which, in the deep infrared, is a mass term for the dual photon.   This is the type of instability that we will look for in the Polyakov models with massless fermions and some related gauge theories. 

We will see that the microscopic symmetries $ {\cal G_M} $ and a topological shift symmetry which arises as a natural consequence of the Callias index theorem  very severely  restrict the types of operators that can be generated. In some circumstances, the infrared theory is quantum critical,  in the sense that there exists no perturbative or nonperturbative operators which may destabilize the masslessness of photons and fermions, and    some such theories become conformal field theories.

\subsection{Callias index theorem and (continuous) topological symmetry}  
In the presence of massless (or light) fermions, the monopoles 
  may  carry fermionic zero modes 
 attached to them \cite{Jackiw:1975fn}.  The number of the fermionic insertions is determined uniquely by the 
  Callias index theorem \cite{Callias:1977kg}, and matter content of the theory. \footnote{ For the 
  relation between the more familiar Atiyah-Singer index theorem and Callias index theorem in QCD-like gauge theories on small $S^1 \times \R^3$, see page.37 of \cite{Shifman:2008ja}.}
Let   ${\cal I}_{\bm \alpha_i}$  denote the  index associated with the monopole with charge ${\bm \alpha_i}$.   
 The typical form of the monopole operator  in the theory with fermions is 
\begin{equation}
e^{-S_0} e^{ \pm i \bm \alpha_i \bm \sigma} O_{\rm fermions}. 
\end{equation}
The  number of   fermion insertions  of each flavor/type, say $\psi_1^a$,   
in $O_{\rm fermions}$, is determined by the  index 
  ${\cal I}_{\bm  \alpha_i}$, by the difference of  the dimensions of the zero energy eigenstates: 
\begin{eqnarray}
{\cal I}_{\bm \alpha_i} = 
( \dim \ker \Dslash_{ \bm \alpha_i } - \dim \ker \overline \Dslash_{ \bm \alpha_i} ) 
\end{eqnarray}  
Here, $\Dslash_{ \bm \alpha_i } = [\sigma_{\mu} (\partial_{\mu} + i A_{\mu})+  \sigma_4 (i\Phi)]_{\bm \alpha_i} $ is the Dirac-like  operator in $d=3$  dimensions in the background  of the 
monopole $\bm \alpha_i$.    In  our conventions, the  $O_{\rm fermions}$ in the monopole operator has only $\psi_a$ insertions, and an anti-monopole operator can only have 
$\bar \psi_a$  insertions. 
\begin{equation}
e^{-S_0} e^{ +  i \bm \alpha_i \bm \sigma} O_{\rm fermions}( \psi ) , \qquad e^{-S_0} e^{ - i \bm \alpha_i \bm \sigma} O_{\rm fermions}( \bar \psi ) 
\end{equation}
This was indeed the reason for the peculiar  spinor 
decomposition (\ref{split}). For an adjoint fermion, the index is equal to 
${\cal I}_{\bm \alpha_i} =2$. In the presence of fundamental fermions, the index is 
${\cal I}_{\bm \alpha_i} =\delta_{i, \hat i}$ where   ${\hat i}$ is the monopole 
that the  zero mode  is localized into.  This is for each flavor  of two component 
Dirac fermion.  Since we have even number of fundamental fermions,  the 
number of fermionic zero mode insertion in  $O_{\rm fermions}$ is always even. 

More precisely,  for fermions in complex representations, 
we have two Dirac-like operators  as seen in (\ref{fl})  and two conjugates,  
\begin{eqnarray}
&& \Dslash^{(1)}  = \overline \sigma_M D_M=  \sigma_{\mu} (\partial_{\mu} + i A_{\mu})+  \sigma_4 (i\Phi), \qquad   \overline \Dslash^{(1)} =  \sigma_M D_M=  \sigma_{\mu} (\partial_{\mu} - i A_{\mu})+  \sigma_4 (i\Phi), \cr \cr
&& \Dslash^{(2)}  = \overline \sigma_M D_M=  \sigma_{\mu} (\partial_{\mu} - i A_{\mu})-  \sigma_4 (i\Phi), \qquad   \overline \Dslash^{(2)} =  \sigma_M D_M=  \sigma_{\mu} (\partial_{\mu} + i A_{\mu}) - \sigma_4 (i\Phi), \qquad \qquad \,
\end{eqnarray}
The total number of fermion zero modes 
associated with a monopole ${\bm \alpha_i}$ is 
 $  n_f(  {\cal I}_{\bm \alpha_i}^{(1)} +  {\cal I}_{\bm \alpha_i}^{(2)} )= 
  2n_f {\cal I}_{\bm \alpha_i} $.

{\bf Symmetry transmutation:} The microscopic Polyakov Lagrangian with massless fermions  has a   global $U(1)_A$   symmetry  given in (\ref{symfun}) and (\ref{symadj})
regardless of whether fermions are in a  real or complex representation. 
Since it is a non-anomalous symmetry, it must be a symmetry of the long distance theory. 
  The $U(1)_A$ transformation,
  \begin{equation}
    \psi \rightarrow  e^{i\beta} \psi , \;\; \;  \bar \psi \rightarrow  
  e^{- i\beta} \bar \psi
  \label{sym}
  \end{equation}
 implies
$ O_{\rm fermions}  \rightarrow e^{i N_f    {\cal I}_{\bm  \alpha_i} \beta } O_{\rm fermions} $.
 Therefore,  the invariance of the monopole operator under (\ref{sym}) necessitates   a 
 continuous  shift   for the dual photons:
\begin{equation}
U(1)_{*} : \; \bm \alpha_i \bm \sigma \rightarrow \bm \alpha_i \bm \sigma - N_f   {\cal I}_{\bm  \alpha_i} \beta 
\label{topsym}
 \end{equation}
Since this symmetry originates from the topological index theorem, we will call it 
a topological shift symmetry, or simply, {\bf  topological symmetry} and 
refer to it as  $U(1)_{*}$.  
Just like the abelian duality transform \cite{Polyakov:1976fu}, 
 the  topological shift symmetry requires going  to sufficiently long distances. 
 In the IR,  the $U(1)_A$ symmetry of the original theory intertwines   with the   shift symmetry
for the dual photons (\ref{U1J}).  This phenomena pervades the physics of  all 
 P$({\cal R})$ theories.  
  
More precisely, recall that in the absence of fermions and monopoles, the free Maxwell   theory 
is dual to a free scalar theory with a continuous shift symmetry $U(1)_{\rm flux}$ (\ref{U1J}).  The presence of monopoles (in the absence of fermions) spoils this symmetry completely. However, 
in the presence of fermions,   the   $U(1)_{*}$  linear combination of the $U(1)_A$ and $U(1)_{\rm flux}$ 
 \begin{equation}
U(1)_{*} : U(1)_A - N_f {\cal I}_{\bm  \alpha_i} U(1)_{\rm flux} 
\label{topsym2}
 \end{equation}
 remains  a true symmetry of the theory. \footnote{If there was no dual photon field to soak-up 
 the phase of the fermionic zero modes,  this would indeed imply that $U(1)_A$ must be anomalous, which is incorrect on $\R^3$.   Compare this with  one flavor QCD on  $\R^4$.
 The instanton vertex  also has two fermion insertion and no extra structure to soak-up the  
 $U(1)_A$ chiral  rotation. Indeed,  there is a chiral anomaly on $\R^4$ and the $U(1)_A$  is anomalous. Only a $\Z_2$ subgroup of it is anomaly-free.  } 
   
  A  continuous  shift symmetry can protect a scalar from acquiring a mass. 
  Since there is only one parameter in 
 the  transformation (\ref{topsym}),  only one  dual photon is protected by the topological symmetry.  At a conceptual level,  this  shows that one gauge degree 
 of freedom remains massless in the IR of  the  P(${\cal R}$)  theory regardless of any other detail, so long as the microscopic theory possesses the $U(1)_A$ symmetry.  
 We may call this phase   deconfined, since a gauge boson remains infinite ranged. Although this is true, it is a crude  characterization.  A more refined categorization of the deconfined phases, which can   distinguish a  free infrared theory (free photon), and 
a strongly or weakly coupled  conformal field theory (CFT)
  is needed, and  will be   discussed.

 \subsection{Revisiting P(adj): Dual scalar  as a Nambu-Goldstone boson} 
 Consider  the $SU(2)$        one flavor P(Adj). (Below is a review and 
 slight refinement  of Affleck et.al. \cite{Affleck:1982as}).  We assume the  long distance gauge 
 structure reduces down to  $U(1)$.   Perturbatively, we have a photon and a neutral fermion, 
 described by 
\begin{equation}
 L= \frac{1}{4 g_3^2} F_{\mu \nu}^2  + i \bar \psi \sigma_{\mu} \partial_{\mu} \psi 
 \end{equation}
 a free field theory. Parity   forbids relevant 
 perturbative operators such as  $\bar \psi \psi$ from being generated 
 \cite{Affleck:1982as}. 
   Nonperturbatively,  there is only one type of elementary monopole (and its anti-monopole.) The index ${\cal I}_{\alpha_1}=2$ for adjoint fermions. Thus,  by (\ref{sym}) and (\ref{topsym}), 
  we have 
 \begin{equation}
   \psi  \rightarrow  e^{i \beta} \psi , \qquad  \sigma \longrightarrow \sigma - N_f {\cal I}_{\alpha_1} \beta = \sigma- 2 \beta
\label{symadj2}.
   \end{equation}
There is   only one combination of the relevant  ${\cal G_M}_{,\rm P(adj)}$    singlet that one can construct, and which gets   induced nonperturbatively:
  \begin{eqnarray}
\Delta L^{\rm non-pert.} =   e^{-S_0} e^{i \sigma} \psi  \psi  +  e^{-S_0} e^{- i\sigma} \;    \bar\psi  \bar \psi   
  \end{eqnarray}
  There is also a large class of   ${\cal G_M}_{,\rm P(adj)}$ singlet, but irrelevant  
  multi-monopole operators of the form  $ ( e^{-
  S_0} e^{i \sigma} \psi  \psi )^k$ where $k$ is some integer. 
The continuous shift symmetry  (\ref{symadj2}) forbids any kind of potential (such as $e^{i\sigma}+{\rm c.c.}$), the mass term  for the dual photon.  
This proves that  the photon must remain massless nonperturbatively. 
Affleck et.al. showed that,  
by expanding 
the $\sigma$ fields around, say,  zero,  the $U(1)_{*}$ symmetry is spontaneously 
broken and the photon is the Nambu-Goldstone boson. 
 The fermion acquires mass  
$\sim e^{-S_0}$
due to $U(1)_{*}$ breaking.   This is the  
 conventional way to have gapless scalars  in a gauge field theory.
For a fuller discussion, see ref.\cite{Affleck:1982as, Unsal:2007vu}.  For $SU(N)$ and 
multi-flavor generalizations,  see \cite{Unsal:2007jx}.  

It is useful to think of the Noether current associated with the symmetry  (\ref{symadj2}) 
 in the $n_f$ flavor    theory.  It is 
\begin{equation}
K_{\mu}= \bar \psi \sigma_{\mu} \psi - n_f {\cal I}_{\alpha_1}  \partial_{\mu} \sigma = \bar \psi \sigma_{\mu} \psi-  n_f {\cal I}_{\alpha_1}   {\cal J}_{\mu}
= j_{\mu} -  n_f {\cal I}_{\alpha_1}   {\cal J}_{\mu}
 \end{equation}
Recall from $\S$.\ref{sec:Pol} that the current associated with $U(1)_{\rm flux}$ satisfies 
 $  {\cal J}_{\mu}=  \partial_{\mu} \sigma = \half  \epsilon_{\mu \nu \rho} F^{\nu \rho} =  F_{\mu}$ 
 where $F_{\mu}$ is  the magnetic field. 
 Using $\partial_{\mu} F_{\mu} = \nabla^2 \sigma= \rho_m(x)$ where $ \rho_m(x)$ is the magnetic charge density, 
  the local current conservation can be re-expressed as 
 \begin{equation}
\partial_{\mu} K_{\mu} = \partial_{\mu} (j_{\mu}  -   n_f {\cal I}_{\alpha_1}   {\cal J}_{\mu}) =0  \Longrightarrow \partial_{\mu} j_{\mu} (x)=   n_f {\cal I}_{\alpha_1}  \rho_m(x)
 \end{equation}
 which implies the conservation of the  $U(1)_{*}$ 
 current as stated in (\ref{topsym2}).  
The final form is  the local version of the  Callias index theorem, which ties the  $U(1)_A$ charge 
with the $U(1)_{\rm flux}$ charge. 
Namely, in the presence of  $n_f$  adjoint fermions,  
\begin{eqnarray}
Q_{*} &=& Q_A -   n_f {\cal I}_{\alpha_1}  Q_{\rm flux} \cr 
&=&N_\psi - N_{\bar \psi} -  n_f {\cal I}_{\alpha_1}   (N_{\rm monopole} -N_{\rm anti-monopoles})
\end{eqnarray}
is a conserved charge, where  $N_X$ counts  the number of the $X$  excitations. 
This means, any    perturbative or non-perturbative interaction vertex in the long distance theory preserves $Q_{*}$.  However, the $U(1)_{*}$ is spontaneously broken by the vacuum, and  
the photon is a Goldstone boson.

{$\bf SU(N)$:} It is also useful to review the $SU(N)$ generalization of this theory since it carries important lessons on the interplay of symmetry and dynamics.  
Due to gauge symmetry breaking down to $U(1)^{N-1}$, 
there exist   $N-1$ photons and $N-1$ massless fermions,  the components along the Cartan subalgebra. 
The infrared Lagrangian in perturbation theory is, therefore, 
\begin{eqnarray}
L^{\rm pert. theory} = \half (\partial  \bm \sigma)^2 +i \bm {\bar \psi} \sigma_{\mu} \partial_{\mu} \bm \psi, 
\qquad \bm \sigma \equiv (\sigma_1, \ldots, \sigma_{N-1}),  \; \; \bm \psi \equiv 
(\psi_1, \ldots, \psi_{N-1}),  
 \end{eqnarray}
 The simplicity of this system relative to the complex representation fermions to be studied in the subsequent section 
is the electric neutrality of the zero mode fermions.   In perturbation theory, 
there are no relevant or marginal operators 
which respect the underlying symmetries of the original theory and which may be generated perturbatively. Thus, the  Gaussian fixed point is stable to all orders in perturbation theory.

However,   there exist a plethora of relevant  nonperturbative (flux) operators. 
The index is  ${\cal I}_{\bm \alpha_i}= 2 $ for all $i=1, \ldots N-1 $. The $N-1$ monopole  operators are   
$e^{-S_0} e^{i \bm \alpha_i \bm \sigma} {\bm \alpha}_i \bm \psi \bm  \alpha_i \bm \psi $ none of which generates a mass term for the dual photons.   Notice that each term is manifestly invariant under the topological 
$U(1)_*$ symmetry (\ref{sym}), (\ref{topsym}). 
In the  $e^{-S_0}$ expansion,  at order $e^{-2S_0}$, 
there are $N-2$ linearly independent relevant operators,   
$e^{-2S_0} e^{i ({\bm \alpha}_j-{\bm \alpha}_{j+1}) \bm  \sigma} $ which get generated. Even though there is no fermion zero mode attached to these topological objects, since they are essentially the bound states of a monopole (with charge $\bm \alpha_i$) and anti-monopole  (with charge $-\bm \alpha_{i \pm1}$),
 their invariance under the $U(1)_{*}$ topological symmetry is also manifest. \footnote{ 
 A  monopole and antimonopole in the presence of massless adjoint fermions interacts logarithmically at large distances in Euclidean $\R^3$, rather than the Coulomb's 
law. (Also see  \cite{PhysRevB.39.8988,Marston:1990bj} for $U(1)$ QED, but one needs to be really careful here.  See formula (\ref{mmint}) and the discussion around it.)
The $\log|x-y|$ marginally binds a monopole into its antimonopole.  The combined state is  magnetically neutral, and cannot lead to Debye screening. (A  monopole-antimonopole pair is  a dipole, and in the long distance, the dipole-dipole interaction is $1/r^3$, hence the absence of the Debye screening.)  In P(adj) with $N\geq 3$, the presence of the 
fermion zero modes also 
 leads to $N-2$ bound states  of a monopole with charge $\bm \alpha_j$ and antimonopole with charge $-\bm \alpha_{j\pm1}$. The combined topological excitation has a nonzero magnetic charge $\bm \alpha_j - \bm \alpha_{j \pm 1}$ and  at large distances interacts via Coulomb potential, $1/r$.
 These excitations are  referred to as magnetic bions \cite{Unsal:2007jx}.    The magnetic bions render $N-2$ varieties of $N-1$ photons massive.
In QCD(adj) on $S^1 \times \R^3$ discussed  in  Ref.\cite{Unsal:2007jx}, due to an extra elementary monopole,  one can form $N-1$ magnetic bions, and the gauge sector is fully gapped. This also has a nice symmetry interpretation. The $U(1)_{*}$  continuous topological shift symmetry turns into a  $(\Z_N)_{*}$
discrete shift symmetry on small $S^1 \times \R^3$. The discrete shift symmetry cannot prohibit  mass  term for scalars. 
} 
Thus, the combined nonperturbative effects up to order $e^{-3S_0}$ is given by 
\begin{eqnarray}
\Delta L^{\rm non-pert.} = 
e^{-S_0} \sum_{j=1}^{N-1} e^{i \bm \alpha_j {\bm \sigma}} \bm \alpha_i \bm \psi  \bm \alpha_i
\bm \psi  + 
e^{-2S_0}  \sum_{j=1}^{N-2} e^{i (\bm \alpha_j-\bm \alpha_{j+1})  {\bm \sigma}} + \; (\rm conjugates) 
\end{eqnarray}
 This renders $N-2$ varieties of the photons massive leaving the one which is protected by the shift symmetry.  As in the $N=2$ case, the $U(1)_{*}$ breaks down spontaneously and there exist only one Goldstone.  The higher order terms in the  $e^{-S_0}$ do  not alter this conclusion. 

This application  shows that the existence of   $U(1)_{*}$ symmetry  provides a 
characterization for the absence of mass gap in gauge sector and the absence of confinement.  
The $U(1)_{*}$ does not imply the absence of monopoles or the irrelevance of monopole operators. And neither the presence of elementary monopoles or magnetically charged bound states of the monopoles implies confinement.

\subsection{Complex representation fermions,  masslessness and quantum  criticality}  
\label{sec:F}
 Let us consider  an $SU(2)$  Yang-Mills noncompact adjoint Higgs system with 
 $N_f= 2n_f$  two component  fundamental Dirac fermions on $\R^3$, the P(F) theory. The theory possess the  symmetries (\ref{master1}).
  As always, we assume the  
$SU(2)$ gauge structure  reduces   down to $U(1)$ at long distances.  
 The off-diagonal gauge degrees of freedom ($W$-bosons) and one component of the fermions in the $SU(2)$ gauge symmetry doublet, and the scalars acquire masses and decouple from the long distance 
physics. 
In perturbation theory, the infrared theory is described  by the 
abelian QED$_3$ action 
\begin{eqnarray}
S^{\rm P(F)}_{\rm pert.}= \int_{\R^3}  \;  \Big[  \frac{1}{4 g_3^2} F^{ 2}_{\mu \nu}  \; + \; 
     i \bar \Psi^a \gamma_{\mu} (\partial_{\mu} + i A_{\mu})   \Psi_a 
    \Big]   
    \label{QED}
\end{eqnarray}
The action possesses an enhanced (accidental)  $SU(2n_f)$ flavor symmetry 
group, and a $U(1)_V$ symmetry which is the  global part of the gauge symmetry.
 This enhancement is expected in perturbation theory, because 
the Higgs scalar acquires mass and disappears from the long distance description. Since the 
 disparity between the gauge-kinetic term and Yukawa term  in (\ref{fl}) was the source of the lower symmetry, and since there are no Yukawa's in the long distance limit, there is an enhanced symmetry in perturbation theory.

The  non-perturbative effects   may in principle be aware of the lower 
symmetry of the high energy theory, and indeed, they are. 
Let us first take $N_f=2$. 
 As in P(Adj),  there is one type of monopole.  The index theorem 
  tells us that for each fundamental flavor,  the monopole has ${\cal I}_{\bm \alpha_1}= 1$ zero mode.  
 There is one  relevant $\G_{\cal M}$  singlet operator  which is induced nonperturbatively:
  \begin{eqnarray}
{\rm Relevant} \;   \G_{\cal M} \; {\rm singlets} :  \;\;\; 
  e^{-S_0} e^{i \sigma} \psi_1 \psi_2  +   e^{-S_0} e^{- i\sigma} \;    
  \bar\psi_1 \bar \psi_2  
  \end{eqnarray}
  The two fermions and the dual photon transform under   $U(1)_{*}$ as 
\begin{eqnarray}
&& U(1)_{*}:  \psi_1 \rightarrow e^{i \beta } \psi_1, \qquad  \psi_2 \rightarrow e^{i \beta } \psi_2, \qquad 
\sigma \longrightarrow \sigma - 2 \beta \; .
\end{eqnarray} 
The continuous shift symmetry forbids any kind of mass term for the dual photon.  In particular, 
it forbids the  $e^{-S_0} (e^{i \sigma} +e^{-i \sigma} )  $ operator.  
Thus, the photon must remain massless nonperturbatively. 

In the multi-flavor case $N_f=2n_f \geq 4$,  the simplest monopole operator  has $2n_f$ insertion 
of the fermionic zero modes, 
$$e^{-S_0} e^{i \sigma} \left[ (\psi_1^1 \psi_2^1)\ldots  (\psi_1^{n_f} \psi_2^{n_f}) +  {\rm permutations} \right]$$
The equality of the number of $\psi_1^a$ insertion with the  $\psi_2^a$ insertion is a consequence of the Callias index theorem and   $U(1)_V$ symmetry, i.e, electric charge  neutrality.   
Making the $SU(n_f)_1 \times SU(n_f)_2  $ symmetry of the monopole operator manifest 
gives  
 \begin{eqnarray}
  \G_{\cal M} \; {\rm singlets} :  \;\;\; 
  e^{-S_0} e^{i \sigma} \det_{a, b} \psi_1^a \psi_2^b  +   e^{-S_0} e^{- i\sigma} \;     
  \det_{a, b}  \bar\psi_1^a \bar \psi_2^b  
  \end{eqnarray}
where $a, b=1, \ldots n_f$ are flavor indices. The invariance of the vertex under 
$U(1)_A$  symmetry 
necessitates the dual photon to transform as 
$\sigma \longrightarrow \sigma - 2n_f \beta$ under  $U(1)_{*}$. 

We identified a  distinction   between the behavior of $N_f=2$  and $N_f \geq 4$ theories. 
In the  $e^{-S_0}$ expansion, the leading   non-perturbatively generated flux operator is classically relevant in the 
$N_f=2$ case, and irrelevant in the $N_f \geq 4$ cases.  Therefore,  the latter class of theories are quantum critical, and will  exhibit enhanced  $SU(2n_f)$ symmetry at  long distance. 
For the $N_f=2$ case, there is one relevant direction and   no enhancement of 
flavor symmetry takes place.

It is again useful to study the Noether currents in the effective long distance theory. Unlike 
P(Adj), there  are two   types of conserved $U(1)$ currents in the Polyakov model with  
$n_f$ complex representation fermions. One is  associated with 
$U(1)_V$ symmetry, and the latter is  a linear combination of  $U(1)_A$  and $U(1)_{\rm flux}$. 
   These are, in the conventions of $\S$.\ref{introcomp},  
 \begin{eqnarray}
&&J_{\mu}= j_{1, \mu} + j_{2, \mu} =  \bar \psi^a_{1} \sigma_{\mu} \psi_{1,a} +  \psi^a_{2} \sigma_{\mu} \bar \psi_{2,a}  \cr  \cr
&&K_{\mu}= j_{1, \mu} - j_{2, \mu} -   2n_f  {\cal I}_{\alpha_1} \partial_{\mu} \sigma = 
 \bar \psi^a_{1} \sigma_{\mu} \psi_{1,a} -    \psi^a_{2} \sigma_{\mu} \bar \psi_{2,a}  -
 2n_f  {\cal I}_{\alpha_1}  \partial_{\mu} \sigma 
 \end{eqnarray} 
The conserved charge  associated with the $U(1)_V$  current $J_{\mu}$ is 
\begin{equation}
(N_{\psi_1} -  N_{\bar \psi_1} )+  (N_{\bar \psi_2} -  N_{\psi_2} )
\end{equation}
and the conserved charge associated with $U(1)_{*}$ is  
\begin{eqnarray}
Q_{*} &=& Q_A -  2 n_f {\cal I}_{\alpha_1}  Q_{\rm flux} \cr
&=& (N_{\psi_1} -  N_{\bar \psi_1} )-  (N_{\bar \psi_2} -  N_{ \psi_2} ) -
 2 n_f {\cal I}_{\alpha_1} (N_{\rm monopole} -N_{\rm anti-monopoles})
\end{eqnarray}
Clearly, these symmetries are in accord with the monopole operators and their zero mode structures.  In fact, the conservation of the $U(1)_{*}$  current, $ \partial_{\mu}K_{\mu} =0$  is the local re-incarnation of the  Callias index theorem. 
We will discuss the infrared limit of these theories  after generalizing the basic essentials 
to $SU(N)$ gauge theory. 

{$\bf SU(N)$:} The difference of long distance physics between $N_f=2$  and $N_f\geq 4$ is 
 not special to the $SU(2)$  P(F) theory.  The infrared limit of   $N\geq 3$   $SU(N)$   
   gauge theory with  $N_f$ massless fermion flavors turns out to be rather similar to the 
     $N_f$ flavor   $SU(2)$ theory, as a consequence of the  non-perturbative dynamics. 
   
We assume    
the gauge structure reduces into  $SU(N) \rightarrow [U(1)]^{N-1}$ at long distances.  In perturbation 
theory, the infrared has 
$N-1$ types of the massless photons, and $2 n_f$    massless fermions. The other 
fields  acquire masses and decouple from the long distance physics.     
There are $N-1$ varieties  of elementary 
monopoles.  Their   Callias indices   are given 
by   ${\cal I}_{\bm \alpha_i} = \delta_{i, 1} = (1, 0, \ldots, 0)_i, \;  i=1, \ldots, N-1 $ where without loss of generality, we assumed that the fermion zero mode is localized into the monopole with charge  $\bm \alpha_1$. 
Thus, the $U(1)_{*}$ shift symmetry reads 
\begin{eqnarray}
&& \bm \alpha_1{\bm \sigma} \rightarrow \bm \alpha_1{\bm \sigma} - (2n_f) \beta ,  \qquad \cr
&&   \bm \alpha_j{\bm \sigma} \rightarrow \bm \alpha_j {\bm \sigma}, \qquad  j=2, \ldots N-1 
 \end{eqnarray}
 The  symmetries do  not forbid  the  $N-2$ types of monopole operators which 
 do not carry any   fermionic  zero modes.  The first monopole  has  $2n_f $ fermion insertions and is    irrelevant for  $2n_f \geq 4$.   
 The  list of all the  flux  operators invariant under the symmetries of the microscopic theory up to order $e^{-2S_0}$ is  
  \begin{eqnarray}
 \;  {\cal G_M}  \; {\rm singlets} : \left\{ \; \;    e^{-S_0} e^{i  \bm \alpha_1 {\bm \sigma}} \det_{a,b} \psi_1^a \psi_2^b,  \; \; 
  e^{-S_0} e^{i  \bm \alpha_2 {\bm \sigma}},  \ldots,     
   e^{-S_0} e^{ i \bm \alpha_{N-1} {\bm \sigma} }    \; \;  \right\}   +   { \rm c.c.} 
  \end{eqnarray}  
Hence, $N-2$ out of $N-1$ photons acquire mass due to relevant  monopole induced effects. 
Thus, the $SU(N)$  P(F) theory undergoes changes in its gauge structure as we consider longer and longer length scales.  The first change is  
 perturbative  $SU(N)  \longrightarrow  [U(1)]^{N-1} $
and  the latter is non-perturbative $[U(1)]^{N-1} \longrightarrow U(1)$ as shown in (\ref{pattern1}). 
The very long distance $U(1)$ theory is quantum critical due to the absence of any relevant 
or marginal perturbations which may destabilize its masslessness. We will comment on the effects of strong (non-compact) gauge fluctuations in the next section.

 Note that  regardless of the value of the rank $N$ in the original gauge theory,  the deep   
 IR of the P(F) theory  always 
 reduces to  an abelian  $U(1)$ QED$_3$  theory with $2n_f$ flavors.  Below, we discuss the long distance limit of  this theory.  
 
\subsection{Conformal field theories (CFTs) at long distances}
  $\bf 2 n_f \geq 4:$
 The $U(1)_{*}$ topological symmetry combined with symmetries such as parity, 
Lorentz  and flavor symmetries forbids  any relevant instability that may occur in the infrared limit of our theory.  The monopole   operators such as   $e^{i \sigma}  $, or   $e^{i \sigma} ({\rm fermion \;  bilinears})  $, where $\sigma$ is the dual of the final  $U(1)$ factor,  are   forbidden.  This means, in the compact continuum QED$_3$ theory obtained as described above, {\bf there are no relevant flux  (monopole)  operators  in the original ``electric"  theory. } 
Thus,  the  non-perturbative  lagrangian is the same as the perturbative one,  
\begin{eqnarray}
S^{\rm P(F)}_{\rm non pert.}= S^{\rm P(F)}_{\rm pert.} \; + \; \ldots  
    \label{QED2}
\end{eqnarray}
where ellipsis stands for  irrelevant perturbations consistent with the microscopic symmetries of the underlying theory. This is 
  QED$_3$  with   charged massless fermions, and with an enhanced (accidental) 
  $SU(2n_f)$   flavor symmetry. 

 The  theory (\ref{QED2})  has no dimensionless coupling constant.  The expansion parameter is 
 $\frac{g_3^2}{k }$ where $k$ is some  euclidean momentum scale.  Thus, 
  perturbative techniques  are  not useful  at low energies. The low energy limit is a  strongly correlated  system  of fermions and gauge fluctuations whose masslessness is protected 
  by $U(1)_{*}$.   A    logical 
  possibility  for the infrared theory is a weakly or strongly  coupled conformal field theory (CFT) depending on the number of flavors. In order to see this, let us calculate the correction to the photon 
  propagator at one loop order in perturbation theory. Partially integrating out fermions 
  produce the non-analytic correction to the gauge kinetic term 
   \begin{equation}
  \frac{1}{g_3^2} F_{\mu \nu}^2 \rightarrow \frac{1}{g_3^2} \left(F_{\mu \nu}^2 + \frac{g_3^2 n_f }{8}  F_{\mu \nu} \frac{1}{\sqrt \Box }  F_{\mu \nu} \right) \; .
\end{equation}
 In the large $n_f$ limit, the higher order effects in perturbation theory are suppressed by powers 
 of $1/n_f$ and the one loop result becomes reliable \cite{PhysRevLett.60.2575}.  The low energy limit is the same as taking $g_3^2$ to $\infty$.  These changes in the photon propagator can be summarized as 
  \begin{equation}
  \frac{g_3^2}{k^2} \underbrace{\longrightarrow}_{\rm one-loop} \frac{g_3^2}{k^2 + \frac{g_3^2}{8} 
   n_f k} \underbrace{\longrightarrow}_{\rm low \; energy}  \frac{8}{n_f k} 
   \label{oneloop}
\end{equation}
 Thus, we are left with a theory without any scale in the IR with gauge boson propagator $\sim \frac{1}{ k}$.  Using the canonical normalization for the gauge kinetic term, 
 the Lagrangian can be expressed as  
 \begin{equation}
 L  \sim F_{\mu \nu} \Box^{-1/2} F_{\mu \nu} +   i \bar \Psi^a \gamma_{\mu} (\partial_{\mu} + i \frac{1}{\sqrt {n_f}} A_{\mu})   \Psi_a 
 \label{deepIR}
 \end{equation}
 with a dimensionless expansion parameter $1/\sqrt {n_f}$. 
   This is a remarkable change in the dynamics.
   
    To appreciate this, let us measure the 
 potential between   two external  electric charges  located  at ${\bf x, y} \in \R^2$. 
The  Coulomb potential between the two test charges is  $V_{\rm Coulomb} ({\bf |x-y|}) 
= \log {\bf |x-y|}$,   in two spatial dimensions,  hence marginally confining. 
   The  non-perturbative dynamics of the pure Polyakov  model  alters this  potential  into a 
 linearly confining one. In the infrared of the theory with massless fundamental fermions, 
 the potential is dictated by   conformal behavior. Thus, 
   \begin{equation} 
V_{\rm non-pert.}({\bf |x-y|}) \sim  \left\{
\begin{array}{ll}
 {\bf |x-y|} & \qquad   {\rm pure \; \;  Polyakov  \; or \; with  \; \; heavy \; fermions}  \cr \cr
 {\bf |x-y|}^{-1}  & \qquad  {\rm with \; \; massless \; \; fundamental \; \;  fermions} , \cr\cr 
\log {\bf |x-y|}  & \qquad  {\rm with \; \; massless \; \; adjoint  \; \;  fermions} , 
\end{array}
\right.
\label{CFT-confine}
 \end{equation}
 In some sense, the long distance behavior of the Polyakov model with massless fermions is more drastic than the Polyakov model {\it per se}.  This  example also shows that  the presence of a single  massless fermion  can completely alter the confining property of the gauge theory! 
 However, the main concept here is not really the presence or absence of a fermionic species. Rather, it is the 
 nature (continuous versus discrete)  of the topological symmetry, as we will discuss in more detail, especially in connection with QCD* theory.
 
The microscopic symmetries of the P(F) theory given in  (\ref{master1}) enhances and 
transmutes into 
  \begin{eqnarray}
{\cal G}_{\rm IR, P(F) } \sim   ( {\rm conformal \; symmetry} )   \times C \times P \times T \times 
U(1)_V \times U(1)_{\rm flux}  \times SU(2n_f)
\cr
\label{master1conf} 
\end{eqnarray}
in the long distances. 
In the $2n_f \geq 4$ cases,  the relevant $U(1)_{*}$ respecting operators also individually respects  $U(1)_A$ and $U(1)_{\rm flux}$.  The $U(1)_A$ is part of  $SU(2n_f)$, and $U(1)_{\rm flux}$ is the symmetry associated with conservation of magnetic flux.   
In the $2n_f=2$ case, only the $U(1)_{*}$ combination is a symmetry.  

 Eq.(\ref{master1conf}) is indeed the symmetry group of the algebraic spin liquid discussed 
in \cite{hermele-2005-72}. 
 The P(F) theory, just like the spin liquids,  undergoes enormous 
space-time and flavor symmetry enhancement. [Compare the long distance symmetries with the 
short distance ones, (\ref{master1}).]
 Interestingly, very different microscopic theories (one is lattice spin system in the $\pi$-flux or  staggered flux state and the other is continuum P(F) theory)   both  flow to the identical long distance interacting  CFT.  \footnote{Recently,  the gauge/string  (AdS/CFT) correspondences 
  are receiving much attention to model QCD-like gauge theories in 4d and lower dimensional condensed matter systems.  Although there is currently no complete matching 
which captures both microscopic and macroscopic aspects of  the 
  most interesting gauge theories (such as the ones appearing  in Nature), 
   it certainly makes sense to model the infrared CFTs or whatever infrared behavior of the strongly coupled system by using a gravitational dual. Such constructions has computational utility at strong coupling.  It may be useful to construct the gravitational duals of the spin liquids.}
    Thus, the multi-flavor QED$_3$ theories  which descend from the Polyakov model  are   generically quantum critical. A recent work discusses  the finite temperature limit of this class of CFTs  \cite{kaul-2008}.

It is not completely clear what occurs for fewer  flavors.  A logical possibility is that the weakly coupled CFT may interpolate into a strongly coupled CFT.  For $2n_f \geq 4$, 
there is some evidence from the large scale lattice studies that no chiral symmetry breaking occurs in this theory \cite{Hands:2002dv}. These lattice simulations of {\it non-compact} 
QED$_3$ are  relevant to our discussion  only because the effect of compactness of the 
gauge boson  in our  theories with $n_f\geq 2$ 
is irrelevant  in the renormalization group sense. 
Also, the inequality in Ref.~\cite{Appelquist:1999hr} suggests that the $SU(2n_f)$ global 
symmetry should be unbroken for $n_f \geq  2$.  Ref.~\cite{Appelquist:1999hr}  also argues that an earlier bound 
for a larger values  ($3 < n_f<4 $)  \cite{PhysRevLett.60.2575}
is an overestimation  of the truncated  Schwinger-Dyson equations.

  $\bf 2 n_f =2 :$ In the $n_f=1$ case, 
the nonperturbative infrared Lagrangian of P(F) is 
\begin{equation}
L^{\rm P(F)}_{\rm non pert.}= L^{\rm P(F)}_{\rm pert} +
   e^{-S_0} e^{i \sigma} \psi_1 \psi_2  +   e^{-S_0} e^{- i\sigma} \;    
  \bar\psi_1 \bar \psi_2  + \; \ldots
  \label{PF}
\end{equation} 
where ellipsis again refer to perturbations such as   $(e^{-S_0} e^{i \sigma} \psi_1 \psi_2)^k$ with $k\geq 2$ which are allowed by symmetries, but irrelevant in the renormalization group sense. 

 In this case, it is not possible to consult the Monte-Carlo studies for the  noncompact lattice QED$_3$, because the  
 effect of compactness is a relevant perturbation of non-compact QED$_3$ dynamics. 
However, it is certain that, due to  topological $U(1)_{*}$ symmetry,  
the photon  remains gapless.  The strong coupling dynamics in the IR combined with  the existence of a  relevant monopole  operator make the 
determination of the long distance  physics hard, and this is left as an open problem.  

To conclude, in $n_f \geq 2$,  the combination of the  topological symmetry and  the
irrelevance of operators which may lead to the breaking of the global symmetries   not only protects  the dual photon (scalar) from acquiring mass,  it  also protects the   fermions.  The mechanism of gaplessness    is different from the Nambu-Goldstone  mechanism. In particular, it relies on unbroken symmetry.  Protection of masslessness due to unbroken symmetry appeared previously  in the context of strongly 
coupled gauge theories, (see chapter 6 of \cite{Peskin:1982mu} for a review, and references therein). More recently,  refinements and generalization  of this idea appeared in condensed matter context as   quantum order  \cite{wen-2002-65, wen-2002-66}. The   appearance of quantum order  in Polyakov model with fermions in new, and  is one of the  main results of this work.

\section{Topology of  adjoint Higgs field and QCD*}
\label{sec:top}

There is a way to trick the Polyakov model with massless fermions,  and get confinement!
In particular, we will present gauge theories which reduce to 
  (\ref{QED}) in perturbation theory, but  are gapped non-perturbatively.

\begin{FIGURE}[t]
{
  \parbox[c]{\textwidth}
  {
  \begin{center}
  \includegraphics[width=6in]{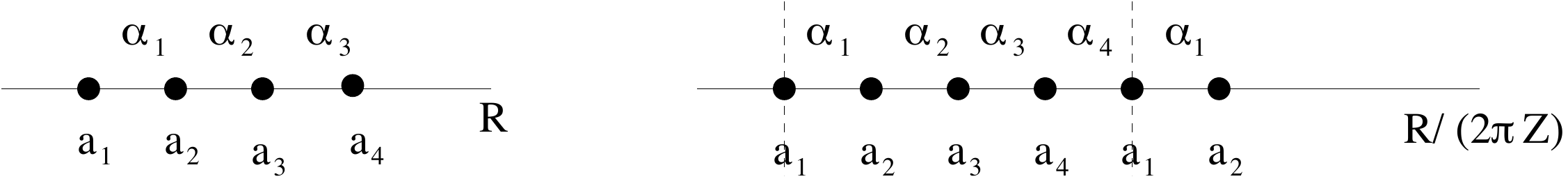}
  \caption
    {%
 The vacuum expectation value of the noncompact versus compact  adjoint Higgs field. Both lead to $SU(4) \rightarrow U(1)^3$ gauge symmetry breaking.  On $\R^3$, when two nearest neighbor 
 eigenvalues become degenerate, the gauge symmetry restore partially and there is an 
 associated elementary  monopole $\bm \alpha_i$. The theory with compact 
 field space topology  has an extra elementary monopole,  $\bm \alpha_4$, which 
``moves in"  from infinity as the    the compactification radius is reduced. When $|a_1({\rm image}) -a_4|$ separation becomes  comparable with the other eigenvalue separation, the extra monopole gains equal fugacity  (or action) with the rest of the monopoles, and contributes equally to the dynamics.   The fundamental distinction between the two models is more pronounced in the presence of massless fermions, as a continuous versus discrete topological symmetry.  
  }
   \end{center}
  }
\label{fig:compact}
}
\end{FIGURE}

 Let us consider the ``identical" looking  action as in (\ref{lagrangian}),  however, alter the topology of the field space into a compact one.  
  Let $\Phi$ be a {\it compact}  adjoint  Higgs field, with a vacuum expectation value   
 $\langle \Phi \rangle ={\rm Diag} (a_1, \ldots a_N)$. These eigenvalues are on the circle (rather than a line) and  $a_N$ is the nearest neighbor of  $a_1$. (See figure \ref{fig:compact}). 
  Naturally enough, this vacuum expectation value will induce the very same gauge symmetry breaking as in the previous sections $SU(N) \rightarrow U(1)^{N-1}$. However, 
  due to the  change in the topology of the field space, there will be an extra elementary monopole 
 other than the ones previously mentioned $\{\bm \alpha_1, \ldots \bm \alpha_{N-1}\} $. The extra monopole stems from the fact that  the eigenvalues  $a_N$ and $a_1$  are now nearest neighbors, and if they become degenerate in real space,  that corresponds to the extra monopole with charge $\bm \alpha_N=-\sum_{i=1}^{N-1} \bm \alpha_i$. The $\bm \alpha_N$ is the affine root of the $SU(N)$ algebra.    This monopole is on the same footing with the rest of the elementary monopoles, in particular, 
 for  $\langle \Phi \rangle$ backgrounds with cyclic   $\Z_N$ symmetry,  the extra monopole  has the same action $ S_{0, N}=S_{0, i} =S_0$, as the rest.

Leaving these secondary  issues aside, let us pose the main question: What did we really change? First, we turned a perturbatively superrenormalizable theory (the case with 
 non-compact adjoint Higgs) into a nonrenormalizable field theory. 
 The latter is in need of a    UV completion. And there indeed exist such UV completions, but these are locally four dimensional QCD-like theories on small $S^1 \times \R^3 $. We assert that,  all the Yang-Mills compact adjoint Higgs theories with or without fermions on $\R^{1,2}$ 
have their UV completion in  QCD-like gauge theories (with judiciously chosen matter content) in 
 small $S^1\times  \R^{1,2}$. Here, however, without concerning ourselves with the  UV completion,  we will only state that the   Yang-Mills {\it compact} Higgs system  on $\R^3$ can  be obtained by adding a  center stabilizing deformation potential  into the 
 YM action  in the small $S^1$ regime, 
   \begin{eqnarray}
 S^{\rm YM^*} &=& S^{\rm YM} + \int_{\R^3 \times S^1} P[\Phi] \cr
&=& \int_{\R^3 \times S^1} \Big[ \frac{1}{4 g^2} \tr F_{MN}^2 + \frac{1}{L^4}
\sum_{n=1}^{[N/2]} a_n |\tr U^n({\bf x})|^2 \Big]
 \end{eqnarray}
 and considering the low energy dynamics of the resultant theory.\footnote{In condensed matter language, this center stabilizing double trace deformation may be viewed as a ${\it frustration}$ of  the Polyakov loop. Without the deformation, in the small $S^1$ regime of YM theory,   
$  \langle \tr U \rangle \neq 0$. At sufficiently large deformation,  
$\langle \tr U \rangle =0 $ even at arbitrarily small $S^1$. A perfect analogy in spin systems is an anti-ferromagnet which upon frustration become a paramagnet, as in (\ref{spin2}). For fruitful applications of this idea into YM and QCD,  see \cite{Unsal:2008ch, Shifman:2008ja}. 
In QCD(adj) formulated on $\R^{2,1} \times S^1$ where $S^1$ is a  spatial circle 
endowed with periodic spin connection for fermions,  these deformations are not necessary, because the quantum
fluctuations (as opposed to thermal fluctuations, which are absent in this setup) prefers a center 
symmetric vacuum \cite{Unsal:2007vu, Unsal:2007jx}.  This theory is  the motivation behind the double trace deformations. }
 Here, $\Phi({\bf x}) \equiv A_{4}({\bf x}) $ is the reduction of the gauge field along the short direction,  which is periodic by  construction.  If we label the holonomy along $S^1$ as  
 $U({\bf x})= P e^{i \int_0^L A_4({\bf x}, x_4) dx_4} \approx  e^{i L \Phi ({\bf x})}  $ where the last 
 equality is correct for smooth fields,  the resulting theory can be brought into the form  (\ref{lagrangian}). 
  Here, $[N/2]$ is the integer part of the half rank of the $SU(N)$ gauge group. 
 The deformation terms with sufficiently large coefficients $a_n$ where $n$ goes all the way to  $[N/2]$ are necessary to have maximal gauge symmetry breaking. 
     Similarly, in theories with fermions, this procedure will produce  a QCD* theory, whose action 
  is 
   \begin{equation}
 S^{\rm QCD^*} = S^{\rm QCD} + \int_{\R^3 \times S^1} P[\Phi] 
 \end{equation}
Study of such deformations  of  YM and QCD-like theories  is relatively recent. The main advantage of this construction is that, some of these deformed theories are solvable in the same sense as the Polyakov model. For example, in YM*, the existence of the mass gap and linear confinement can be shown analytically  \cite{Unsal:2008ch}, despite being locally four dimensional. 

The relevance of  this class of theories to our discussion is that, in perturbation theory, the long distance limits of these theories are  indistinguishable from the appropriate Polyakov models, 
and reduce to the 
$[U(1)]^{N-1}$ QED$_3$ on $\R^3$. Thus, they constitute an alternative way to embed compact QED$_3$ into a continuum gauge theory,  different from Polyakov's original constructions \cite{Polyakov:1976fu}.  Remarkably, the 
non-perturbative aspects of some of these theories are  opposite of the Polyakov model with massless fermions.
Their gauge sectors are  gapped as shown in   the  pattern (\ref{pattern2}) for moderate 
numbers of flavors.

\subsection{Discrete topological  symmetry and mass gap}    
 How can such a ``small" change in the topology of the field space  alter  the IR properties so drastically?  The simplest reason is, as always, through symmetries.  As explained above, the compact adjoint Higgs theories descend from   locally 4d QCD-like theories. 
As it is well known,
 there are chiral anomalies in locally $d=4$ dimensional theories, and since anomalies are  a short distance property, it will clearly distinguish a theory whose base space is  $\R^3$  from another one whose base space is secretly $S^1 \times \R^3$ (even if its  lagrangian is expressed on  $\R^3$.) 
 
 Thus, the true symmetry structure of the  P$({\cal R})$ theory must be different from 
 the QCD$({\cal R})$*.   In, for example, one flavor theories, the $U(1)_A$ symmetry of the  
  P$({\cal R})$ theories is replaced by $\Z_{2h}$ discrete chiral symmetry of locally four dimensional theory.  Here, 
   $h=1$  for a fundamental  and $h=N $ for adjoint fermion. The Callias index theorem  is still valid in the formulation  on small $S^1 \times \R^3$, and its precise relation to the Atiyah-Singer index theorem is  well understood \cite{Shifman:2008ja}.  In the presence of massless fermions, 
the     $\Z_{2h}$  which is the discrete chiral symmetry of the microscopic theory,  intertwines with the  $\Z_{h}$  discrete subgroup of the  $U(1)_{\rm flux}$, schematically as   
\begin{equation}
\psi \rightarrow e^{i \frac{2 \pi}{2h}} \psi, \qquad \sigma \rightarrow \sigma - \frac{2 \pi}{h}
\end{equation}
such that the monopole operators (e.g. $e^{i \sigma} \psi \psi$)
remains invariant. Clearly, this discrete symmetry does not forbid operators such as 
$e^{i h\sigma}$ from being generated, but forbids $e^{i h' \sigma}$ if $h' \neq 0 \; ({\rm mod}\;  h)$.

 The topological $U(1)_{*}$  symmetries of  P$({\cal R})$ theories  
  reduce   into a discrete   symmetry in  QCD$({\cal R})^*$,
   \begin{equation}
\underbrace{ U(1)_{*}}_{{\rm non-compact\;  Higgs \; or \;  P}({\cal R})} \longrightarrow \underbrace{ (\Z_{h})_* }_{{ \rm compact \;  Higgs \; or \; QCD}({\cal R})^* }\; .
 \end{equation} 
 We identified  the fundamental  distinction between non-compact and compact adjoint Higgs systems as a change   in their  microscopic, and consequently, topological symmetry:

 As emphasized  in the discussion of P(${\cal R}$),   $U(1)_{*}$ continuous shift symmetry is 
 able to prohibit mass term for one variety of the dual photon. On the other hand, the 
 $(\Z_{h})_{*}$  symmetry which is  
a { \bf discrete topological symmetry} 
is incapable of  forbidding a  mass term for the dual photon.\footnote{
The discrete topological shift symmetry has a representation dependence. 
It is    $\Z_N$ for adjoint, $\Z_{N+2}$ for symmetric, 
$\Z_{N-2}$ for anti-symmetric and  trivial group $\Z_{1}$ for fundamental fermions. These  are also valid for multi-flavor cases.}
 The discrete symmetries can at best postpone the emergence of the mass 
 term in the $e^{-S_0}$ expansion \cite{Unsal:2007vu, Unsal:2007jx}, but can never forbid it. Thus, 
 in the theory with compact adjoint Higgs field,  
  there is no {\it symmetry} reason    for the photon to remain massless and a mass term is generated. 

At this stage we are conceptually done. But in order to come to a full circle with the first paragraph of the  ($\S$.\ref{sec:top}) which had an emphasis on the topological structure of the 
 field space, let us discuss  an example, given in  \cite{Shifman:2008ja}.
 
 \subsection{Application: QCD(F)* with $n_f=1$}

Consider the  analog of  the gauge theory in  ($\S$.\ref{sec:F}), let $2n_f=2$.  
 Due to the change in topology of the field space, there are two types of elementary monopoles.  
 Their magnetic and topological charges
 $ \left( \int_{S^2} F, \;  \int_{\R^3 \times S^1} F \widetilde F\right)$
 are given by 
 ${\cal M}_1: ( +1, \half), \; {\cal M}_2 :(-1, \half)$ for monopoles,  and 
 $\overline{\cal M}_1: ( -1, +\half),\;  \overline{\cal M}_2 :(+1, -\half)$
for   the  anti-monopoles.      The Callias index for 
 the monopole with quantum number  $( +1, \half)$ is one and the index for the  $(-1, \half)$ monopole is zero. Note that the product  ${\cal M}_1 {\cal M}_2$ is the four dimensional instanton
 vertex and the monopoles can be viewed as constituents of the 4d instanton. The zero modes localizes into one of the constituent monopoles following the ``Higgs regime" criteria in the statement of Callias's theorem \cite{Callias:1977kg}. For a nice lattice realization of the localization property,  see Bruckmann et. al. \cite{Bruckmann:2003ag}.  The monopole operators are  
 \begin{eqnarray}
&& {\cal M}_1(x) = e^{-S_0} e^{i { \sigma}} \psi_1 \psi_2,  \qquad  
 {\cal M}_2(x) = e^{-S_0} e^{- i \sigma} \cr 
 && \overline {\cal M}_1(x) = e^{-S_0} e^{- i \sigma} \bar \psi_1 \bar \psi_2,  \qquad  
 \overline {\cal M}_2(x) = e^{-S_0} e^{+  i \sigma}
 \label{monopole}
   \end{eqnarray}
which only respects  $U(1)_V \times (\Z_2)_A$ symmetries of the microscopic QCD(F)* theory. 
 Unlike the case with non-compact adjoint Higgs fields (\ref{PF}), the dynamics and symmetries 
of the compact Higgs theory admits a  relevant monopole operator without a fermion zero mode insertion:
\begin{eqnarray}
\Delta L^{\rm non pert. }  \sim 
       e^{-S_0}  \cos \sigma +  e^{-S_0} e^{i \sigma} \psi_1 \psi_2 + e^{- i \sigma} \bar \psi_1 \bar \psi_2   \qquad 
    \label{cQED2}
\end{eqnarray}
  The mass for the dual scalar  
is $\sim e^{-S_0/2}$  and is  there 
due to the extra    $ {\cal M}_2(x) +   \overline {\cal M}_2(x) $ 
monopole effect $e^{-S_0} \cos \sigma$. This  potential  pins the scalar at the bottom of the potential. 
 Expanding  $ {\cal M}_1(x) + \overline {\cal M}_1(x)$ at the minimum of $\sigma$ yields $e^{-S_0}  ( \psi_1 \psi_2 + \bar \psi_1 \bar \psi_2)$, a mass for the fermion proportional to 
$e ^{-S_0}$, much smaller than the photon mass. Thus, 
  the dynamical patterns of the theory is 
  \begin{eqnarray}
SU(2) \underbrace{\longrightarrow}_{\rm  Higgsing}  U(1)
\underbrace{\longrightarrow}_{\rm nonperturbative}  {\rm no \; massless \; modes}\; 
\label{pattern3}
  \end{eqnarray}
  which is a special case of (\ref{pattern2}).
For a fuller discussion of  one-flavor QCD-like theories with two index representation fermions, we refer the reader to a joint work  with M. Shifman \cite{Shifman:2008ja}.

Note that the important conceptual distinction relative to the P(F) theory discussed in 
$\S$.\ref{sec:F} is the absence of $U(1)_{*}$ symmetry in the QCD(F)*. In P(F), $U(1)_{*}$
forbids the appearance   of all the flux operators without fermion zero mode insertions, such as 
$(e^{-S_0} e^{i \sigma})^k$ for any integer $k$. In  QCD(F)*,  such operators are allowed by symmetries. 

A consequence of the presence versus absence of a continuous topological symmetry is reflected in the interactions between topological excitations. 
 In P(F) on Euclidean $\R^3$, the long distance interactions of monopoles
with anti-monopoles are necessarily logarithmic, whereas in QCD(F)*,  the ${\cal M}_1(x)$ and 
$\overline{\cal M}_1(y)$ interaction is logarithmic, but ${\cal M}_2(x)$ and $\overline{\cal M}_2(y)$ interacts according to  Coulomb's law as it can be seen by inspecting the leading connected correlator of the monopole operators: 
\begin{eqnarray}
&&V_{1 \bar 1} (x-y)= - \log  \langle {\cal M}_1(x)  \overline{\cal M}_1(y) \rangle  \sim 
 4 \log |x-y|  - \frac{1}{|x-y|}, \cr 
&&V_{2 \bar 2} (x-y)= - \log  \langle {\cal M}_2(x)  \overline{\cal M}_2(y) \rangle  \sim 
  - \frac{1}{|x-y|},
  \label{mmint}   
 \end{eqnarray}
The   $ {\cal M}_2(x), \overline{\cal M}_2(y) $ type  monopoles in QCD(F)* are sufficient to have the usual Debye mechanism, and generate a mass gap for the dual photon. 

\subsection{Remark on accidental continuous topological  symmetry}
\label{accidental}
Evidently, the  presence  of a discrete    $(\Z_{h})_{*}$  topological symmetry is a 
necessary  criteria  for the the presence of  a mass gap in the gauge sector. If a mass term 
for dual photon is not protected by a symmetry, surely, it will get generated.  However, it is also possible that a term allowed by all the symmetries may be irrelevant in the renormalization group 
sense. Thus, the presence of discrete topological symmetry is not sufficient  to conclude that the theory  has a  mass gap and confines. 

Consider the QCD(F)*,  a theory defined on small 
$S^1 \times \R^3$ by construction,    as a function of the  number of flavors.
Assume the number of flavor is large, but not very large so that the four dimensional coupling at the compactification scale is small. 
Indeed, a monopole operator is allowed, and hence is generated. However, a monopole operator may become irrelevant 
if there are a  sufficiently large number of flavors. The classical scaling dimension of the monopole fugacity is $+3$.  
The presence of the massless fermions alters the  quantum scaling dimension for the monopole operator in a significant  way   for  large numbers of flavors $(\sim n_f)$ as shown in  \cite{Borokhov:2002ib}. The continuum analysis  for such QCD(F)*  mimics  the analysis of 
 Hermele et.al.  for lattice QED$_3$ at large $n_f$  \cite{hermele-2004-70}. 
 In both cases,    the  monopole  operator does  scale down to zero at long distances \cite{hermele-2004-70} 
due to large scaling dimension, showing the irrelevance of monopoles and emergence of an accidental $U(1)_{\rm flux}$ symmetry associated with the conservation of gauge flux. 
  Strictly speaking,
  there are some important differences between  lattice QED$_3$ and 
 QCD(F)* to be explained after the discussion of spin liquids.  However, those are immaterial for  the above argument. Thus, in QCD(F)* theory, there must be a critical window for the number of  flavors for  which the  theory is a three dimensional interacting conformal field theory.  It is desirable to understand  the relation between these fixed points and the  
   perturbative  Banks-Zaks fixed point  \cite{Banks:1981nn}. Plausibly, they may be smoothly connected within QCD(F)*. 

\section{Compact lattice QED$_3$ and   $U(1)$  spin liquids}

We have arrived at  a very interesting situation.   There are at least two ways to obtain  
``compact QED with fermions" in $d=3$ dimensions by using {\bf  continuum} field theories. We referred to the theories with  non-compact adjoint Higgs  as P($\cal R$) and the one 
 with compact adjoint Higgs fields as  QCD($\cal R$)*.  
In both case, 
 the  resultant QED$_3$ is  compact by necessity,    because both  are 
 realized via gauge symmetry breaking down to the (compact) maximal torus  $ [U(1)]^{N-1} $.

 It is well-known that pure compact QED$_3$  confines 
 even at arbitrarily weak coupling.   A  controversial question 
 is     what happens to   confinement if one introduces massless fermions.  
  This question is of practical  importance in the context of the stability of the $U(1)$-spin liquids in two dimensions,  a phase which {\it may be} neighbor with the $d$-wave superconducting  phase  in cuprates. \footnote{It is not certain that spin liquids play a role in cuprates.  However, the question of whether doping a spin liquid by charge generates a $d$-wave superconductor is sensible and interesting, and its answer may give insights  into the structure 
  of the  pseudo-gap regime.    }   
    Regardless of the relevance of spin liquids for  cuprates, 
   the stability of the spin liquid is associated with the concept of 
  fractionalization,  which does not arise in any naive way from a collection of electrons, but which may exist due to strong-correlation physics. Therefore, this is a conceptually  interesting and experimentally relevant question.  
 Ref.\cite{sachdev-2002-298, herbut-2003-91,herbut-2003-68} argued that the monopole effects always render the  $U(1)$ spin liquids unstable.  Ref.\cite{hermele-2004-70} showed that there are at least some spin liquids, with gapless fermions and $U(1)$ gauge fluctuations.   These works refers to a particular ``3d lattice QED with massless fermions", with a specific set of microscopic symmetries (sometimes called projective symmetry group (PSG)). 
In the large $n_f$ limit,    Ref.\cite{hermele-2004-70} exhibits by relying on the microscopic symmetries of the lattice theory  and a sophisticated  RG analysis which addresses the light electric and magnetic degrees of freedom simultaneously that 
there are no relevant perturbative or non-perturbative instabilities which may render the photon and fermions massive. 
 
Our  work   shows that the  compact QED$_3$ with fermions may arise 
 in at least two different ways  as in  (\ref{pattern1}) and (\ref{pattern2}),  via non-compact  versus compact adjoint Higgs field. (Moreover, it can also arise from a compact lattice formulation.) 
 The change in the topological structure of the field space  produces
 drastically distinct physics in the IR,  gapless versus gapped gauge sectors in some cases. Thus, 
the  question of the  
   presence or absence of a defonfined phase in compact QED$_3$  in  the {\it  continuum} 
   formulation   is an 
  {\bf ill-defined} question  unless one states the symmetries of the cut-off scale (microscopic)   theories clearly.   (The importance  of symmetries is also emphasized  in  lattice formulations 
   Ref.\cite{hermele-2004-70}. )

The analysis of   Ref.\cite{hermele-2004-70} carefully incorporates  all possible symmetry singlet operators that can be generated perturbatively,  or nonperturbatively via  flux (monopole) operators, in a continuum language, by remaining  loyal  to the symmetries of the microscopic  theory.  This is a basic  principle in any effective field theory construction as stated in $\S$.\ref{setup}, either in  the  continuum limit of lattice gauge theory or the long distance description of a 
gauge theory in which gauge structure changes over length scales. 
By a  careful renormalization group analysis,   Ref.\cite{hermele-2004-70}  shows that 
in the large $n_f$ limit, the
quantum effects turn the monopole operator, which has engineering dimension 
$ +3 $, into an irrelevant operator. The essence of this argument, is that at the IR fixed point, 
the quantum scaling dimension for the monopole operator is large $\sim n_f$  
\cite{Borokhov:2002ib}  and forces  the  monopole operator to scale down to zero at long distances.
The irrelevance of    monopoles is the same as conservation of magnetic flux, and there is an emergent topological   $U(1)_{\rm flux}$ symmetry which characterizes the deconfined nature of this fixed point.  (For the details, see   Ref.\cite{hermele-2004-70}.)

In our analysis of continuum QED$_3$ which  descends from the Polyakov model P(F), 
we did not need such a renormalization group analysis   to show the irrelevance of flux (monopole)  operators  such as $e^{-S_0} e^{i q  \sigma}$ with   $q\geq 1$  
because  they are forbidden to begin with, due to $U(1)_{*}$  topological  symmetry. 
Since this symmetry is independent  of the rank $N$ and the number of flavors $ n_f $, 
the assertion that P(F) theory is always in the deconfined  phase 
did not require a large $n_f$ limit either.

In the next sections, we will discuss whether P(F) theory or QCD(F)* theory has anything to do with the  $U(1)$ compact lattice QED$_3$ with massless fermions. The lattice theory of interest is the one which arise in the $SU(n_f)$ spin systems, which we review next.

 \subsection{From $SU(n_f)$ quantum spin model  to lattice QED$_3$}
 It is useful to  briefly review the route from the spin models to lattice QED$_3$ with massless fermions, and identify the symmetries carefully. \footnote{This section is a  review of known results in quantum spin systems, 
 see  \cite{lee-2004}, and references therein. }
The Hamiltonian of  a  $d=2$ dimensional spin model on a square lattice is 
given by 
\begin{eqnarray}
 H=  
    && J \sum_{\langle \bm  r,  \bm r' \rangle}  \tr \left[ {\bm S}({\bm r}). {\bm  S}({\bm r'}) \right] 
    + \ldots  \cr
    \equiv && 
 J \sum_{a=1}^{\rm dim(adj)} 
  \sum_{\langle  \bm r, \bm r' \rangle}  S_{\bm r}^a S_{\bm r'}^a  + \ldots 
\label{spin}
   \end{eqnarray} 
where  $ J >0 $ is the antiferromagnetic exchange, and ellipsis are higher order terms which 
may ease the frustration of magnetic order.   This term may be due to
 geometric frustration or some other microscopic mechanism. 
 Here, $\bm r, \bm r' $ are  points on a two dimensional (square) lattice and    ${\langle \bm r, \bm  r' \rangle}$ indicates the nearest neighbor interactions.  The hamiltonian has a global $SU(n_f)_D$ spin rotation symmetry group acting by conjugation 
\begin{equation}
{\bm S}({\bm r}) \rightarrow U {\bm S}({\bm r})U^{\dagger}, \qquad  U \in SU(n_f)_D \; .
\label{globalSUnf}
\end{equation}
The subscript $D$ stands for diagonal, due to reasons to be explained in $\S$ \ref{sec:lat}. The ellipsis are assumed to be singlets under the $SU(n_f)_D$ symmetry and the other symmetries of the lattice.  

 The description of an ordered phase in terms of the mean field approximation is well known. 
 A  more  non-trivial aspect in higher dimensional systems   is whether 
 the mean field approach can be usefully applied to a  phase which  refuses to  order.  
 The answer  to this question is relatively recent \cite{PhysRevB.37.580, PhysRevB.37.3774 }, and  eventually leads to the emergence of  gauge structure (and 2+1 dimensional gauge theories)  in  spin systems in two spatial dimensions.  A microscopic   Hamiltonian which may have a non-magnetic  ground state is  a double-trace deformation of (\ref{spin}) 
  \begin{eqnarray}
 H=  
    && \sum_{\langle \bm  r,  \bm r' \rangle} \left[ J  \tr \left[ {\bm S}({\bm r}). {\bm  S}({\bm r'}) \right]  + \frac{J'}{n_f}  (\tr \left[ {\bm S}({\bm r}). {\bm  S}({\bm r'}) \right] )^2 \right]
   \label{spin2}
   \end{eqnarray} 
 For    sufficiently large  positive  $J'$,    despite the  leading anti-ferromagnetic term,  
 no long range magnetic order will appear.  The double trace deformation is same as  
  frustration  for the spin order parameter.

To see this,  the local spin operators  $ {\bm S}_{\bm r}$  are expressed  as a local 
 composite of the fermionic  spinon operators  $f_{\bm r, \beta}$ 
\begin{equation}
 S_{\bm r}^{a} (\bm r)  = f^{\dagger}_{\bm r, \alpha}  T^a_{\alpha \beta}   f_{\bm r, \beta},  \qquad {\rm or } \;\; 
 {\bm S}_{\alpha \beta} = (S_{\bm r}^{a}T^a)_{\alpha \beta}  = 
 f^{\dagger}_{\bm r, \alpha}    f_{\bm r, \beta} - \frac{1}{2n_f} \delta_{\alpha \beta} 
\label{spinon}
\end{equation} 
Supplemented with the constraint that  each site must have occupation number $n_f/2$ (with $n_f$ even), 
\begin{equation}
\sum_{\alpha=1}^{n_f} f^{\dagger}_{\bm r, \alpha}   f_{\bm r, \alpha} =n_f/2 \,, 
\label{constraint} 
\end{equation}
this is an  exact description  of the original spin Hamiltonian. 
    This procedure of breaking the  spin into two fermionic 
spinons is called   slave fermion mean field theory, and (\ref{spinon}) should be viewed as the definition of lattice spinons,   $f_{\bm r, \alpha}$ . The spinons obey canonical anti-commutation relations,  $\{   f_{\bm r, \alpha},   f^{\dagger}_{\bm r', \alpha'} \}  = \delta_{ \bm r r', \alpha 
\alpha'}$ and zero for all other  anti-commutators.  
Clearly, the Hilbert space of the theory without the constraint is vastly larger.

There is an apparent  gauge redundancy 
$ f_{\bm r, \alpha} \longrightarrow e^{i \theta({\bm r}) }  f_{\bm r, \alpha}$
 built-in  the definition of the spinon  operator. 
 The local constraint (\ref{constraint})
guarantees that the 
quartic Hamiltonian  in terms of the spinon operators is same as the original Hamiltonian in terms of spin operators. 
Exploiting the  gauge redundancy provides the  connection between 
purely bosonic spin models and lattice theories with  gauge fluctuations and  fermions.

The spin  Hamiltonian (\ref{spin}) in terms of the spinon operators is  quartic. The $U(1)$ 
lattice QED$_3$ arises in describing the fluctuations of this  system around the  $\pi$-flux ($\pi$F), and the staggered flux (sF) state  \cite{PhysRevB.37.3774}. Here, we only review  the $\pi$-flux state.  Let a  mean field ansatz be denoted by 
\begin{equation}
\overline \chi_{\bm r \bm r'}= \langle f^{\dagger}_{\alpha}(\bm r) f_{\alpha}(\bm r') \rangle \; . 
\end{equation}
 The $\pi$-flux state is the configuration of $\overline \chi$ with flux  $\pi$    through each plaquette on the square lattice, 
 \begin{equation}
 \prod_{\partial p} \overline \chi [\partial p ]=e^{i \pi} =-1
 \end{equation}
  where $p$ denotes an elementary plaquette  and $\partial p $ is the oriented boundary.  
It is clear  that 
  $\chi_{\bm r \bm r'}$ transforms  gauge covariantly,    as a  connection on the lattice.  For low energy considerations, only the phase fluctuations 
  of the ansatz are important. 
 Hence, the terms in Hamiltonian  incorporating the fluctuations and spinon hopping term 
  takes the form 
\begin{equation}
H \sim   J \sum_{\langle \bm  r,  \bm r' \rangle} \bar \chi_{\bf r' \bf r}  f^{\dagger}_{\bm r, \alpha}  
e^{i a_{\bm r, \bm r'}}
 f_{\bm r', \alpha} + \rm h.c.
 \label{latticeQED3}
\end{equation}
which is the fermionic terms in  lattice QED$_3$ \cite{PhysRevB.37.3774}.
   Even though the Maxwell term 
is not present above, it will be produced by  the renormalization group, when one integrates out a thin momentum-shell of  fermions. Hence, we can  add it the the above Hamiltonian. 
 \footnote{
The reader familiar with the  staggered fermions 
 (or Kogut-Susskind fermions)  in lattice QCD will 
realize immediately that the spinons are the analogs of  the staggered fermions  \cite{Kogut:1974ag},  and by construction, we are guaranteed to get a relativistic dispersion relations, and  Lorentz invariance (in a naive continuum limit.) The Dirac algebra and spinors of the continuum theory translates into the $\pi$-flux relation and Grassmann valued operators 
in the (reverse) Kogut-Susskind
construction.}  
The resulting theory is {\it compact} lattice QED$_3$ theory with minimally coupled 
fermionic matter. 

The QED$_3$ also appears in the  more phenomenological proposal of Franz et.al.  \cite{franz-2001-87, franz-2002-66}  and \cite{herbut-2002-66} within  the phase fluctuation model in order to describe the pseudo-gap region of cuprate superconductors. 
 The relation between this approach and    the more microscopic spin liquid approach 
 to the underdoped cuprate superconductors, and in particular,  a relation between the   lattice spinons and nodal quasi-particles is currently not clear.

 \subsection{Reverse engineering of  lattice spinons and  twisting}
 \label{sec:lat}
 
 It is useful to  understand the relation between the symmetries of the compact lattice   QED$_3$ (\ref{latticeQED3}) and continuum QED$_3$ with Lagrangian (\ref{QED}). In particular, 
 considering the important role played by  $U(1)_{A}$ symmetry and Callias  index  theorem in 
 the Polyakov model,   it is desirable to understand whether  an analog of these may arise   in the lattice formulations.  
 
 For ease of presentation,  we relabel   the $2n_f$ fermionic continuum fields   as 
 \begin{eqnarray}
 \{ \psi_{1, a} \; ,\;  \bar \psi_{2,a} \} \rightarrow   \{ \lambda_{1, a} \; , \;   \lambda_{2,a} \} \equiv 
 \{ \lambda_{1}, \  \lambda_{2}, \ldots, \lambda_{2n_f}  \}, \qquad a=1, \ldots , n_f
 \end{eqnarray}
 where Lorentz indices are suppressed. 
The continuum Lagrangian in terms of $\lambda$ fields reads 
 \begin{eqnarray}
{\cal L} =   \frac{1}{4 g_3^2} F^{ 2}_{\mu \nu}  \; + \; 
      \sum_{b=1}^{2n_f}  \>\> i \bar \lambda^b \sigma_{\mu} (\partial_{\mu} + i A_{\mu})   \lambda_b 
        \label{conQED}
\end{eqnarray}
The continuum theory has an  $U(1)_V \times SU(2n_f)$ global symmetry, where 
$U(1)_V$ is the global part of gauge symmetry  and $SU(2n_f)$ is a global flavor symmetry. 
 
In the Polyakov model embedding or ``regularization"  of the compact version of this theory, only 
\begin{equation}
  U(1)_V \times SU(n_f)_1 \times SU(n_f)_2 \times U(1)_A 
     \; \subset\; 
 U(1)_V \times SU(2n_f)
 \end{equation}
is  present, where we loosely view the inverse $W$-boson mass as the  lattice spacing. 
 
 Let us now  reverse engineer the lattice QED$_3$ theory starting with continuum formulation.  This will be useful in understanding   what the lattice symmetries mean in the continuum and ease the comparison with Polyakov's  model.  
Consider  continuum QED$_3$ theory in  Hamiltonian formulation on $\R^{1,2}$ and latticize $\R^2$.    Let us consider the 
 $SU(2) \times SU(n_f)_D $  subgroup of the $SU(2n_f)$  flavor symmetry. Since we are in the Hamiltonian  formulation, we split the Lorentz symmetry into $SO(2)$ and continuous time translations.     The fermions are  in two dimensional spinor representation of $SO(2)$, two dimensional spinor representation of $ SU(2)$ and in the fundamental representation of $SU(n_f)_D$.  Now, we wish to discuss a well 
 defined procedure, called {\it twisting},  which intertwines the Lorentz and flavor symmetry such that  the continuum spinors are mapped into Grassmann valued operators residing on the lattice  sites (the  lattice spinons). 
 In spin systems, the $SU(n_f)_D$ corresponds to 
 the  global rotation symmetry (\ref{globalSUnf}) of the spin. It is also 
the diagonal subgroup  of the 
 $SU(n_f)_1 \times SU(n_f)_2$ decomposition which appeared in the Polyakov model. 
 The    $SU(n_f)_D$  will have no impact in our discussion, so we suppress it.
  
    The fermion 
 $\lambda_{ \alpha, a }$ transforms as  $\lambda \rightarrow O \lambda U^{\dagger}$ under 
 $ O \in SO(2)_L$  
 and $U \in SU(2)$ flavor. We can write every two by two matrix such as  
 $\lambda_{\alpha, a}$ in a 
 basis spanned by the identity and the Pauli matrices  $(1, \sigma_{\mu}, \sigma_{\mu \nu}= i \half \epsilon_{\mu \nu} \sigma_{\mu}\sigma_{\nu}  )$. 
Thus,   
\begin{equation}
\lambda_{\alpha, a} =  (f 1 +  f_{\mu} \sigma_{\mu} + \half f_{\mu \nu}  \sigma_{\mu \nu})_{ \alpha,a}  \qquad \alpha=1,2, \; \; a=1, 2
\label{form}
\end{equation}
This is to say that under the diagonal 
 $SO(2)_D = {\rm Diag} (SO(2)_L \times SU(2)) $ subgroup, the spinor becomes a collection 
 of $p$-forms, one scalar, one vector and one two  form anti-symmetric tensor, which we label
 as  $(f, f_{\mu}, f_{\mu \nu}) $. On the lattice, a $p$-form is naturally associated with a $p$-cell,  zero form with  sites, one form with links, and two form with faces. This  twist  is also sometimes referred to as the   ``Dirac-K\"ahler"  construction in lattice gauge theory  \footnote{This type of decomposition is one of the  cornerstone of the  recent progress in   supersymmetric lattices, see for example,  \cite{Kaplan:2003uh, Catterall:2005eh}.} and is known to be equivalent to staggered fermions. We can map these fermions onto a lattice with half the spacing.  
 The mapping takes the single component Grassmanns $f,  f_1, f_2, f_{12}$ onto the sites 
 $(0,0), (1, 0) , (0,1), (1,1)$, in a unit cell, respectively.  
  The new lattice repeats itself in amounts $(2,0)$ and $(0,2)$ in the $x$ and $y$ directions. 
The twisting  procedure is the reverse engineering of the appendix of   Ref.\cite{hermele-2004-70}. To see this, rewrite (\ref{form}) in the component language:
  \begin{equation}
(\lambda_{ \alpha, a}) =  \left( \begin{array}{cc}
    f+ f_{12} & f_1+ if_2 \cr
    f_1- if_2  &   f-  f_{12} 
\end{array} \right)
\label{form2}
\end{equation}
This is indeed the relation between the lattice spinons and continuum spinors given in 
Ref.\cite{hermele-2004-70} modulo a minor renaming of the fields.  \footnote{This is also the reason why fields that transform in a single valued representation of the lattice point group 
 symmetry 
 maps into the double valued spinor representations under  the continuum Lorentz symmetry.   
 This clearly does not make any sense without the twisting idea, which mixes Lorentz symmetry and some global symmetry. 
This is in fact a  recurring  and fruitful  theme in diverse fields 
of theoretical physics. It  appeared   initially in staggered (Kogut-Susskind) fermions  \cite{Kogut:1974ag},  and most  recently in supersymmetric lattices constructions 
\cite{Kaplan:2003uh, Catterall:2005eh}.  
 It also  arises naturally in  ``topologically'' twisted version of the 
 supersymmetric theories, where under the diagonal subgroup of space-time  and some flavor symmetry, the spinors decompose as $p$-forms, single valued representations \cite{Witten:1988ze}.
 Apparently, such  structures are also ubiquitous in spin systems, in particular, the $\pi$-flux 
 and staggered flux  states  \cite{PhysRevB.37.3774}.  
 }

The discrete rotational symmetries of the QED$_3$ lattice action discussed in Ref.\cite{hermele-2004-70} are in fact the subgroup of  $ G_{\rm discrete} \subset  SO(2)_D = {\rm Diag} (SO(2)_L \times SU(2)) $. In the continuum, when the $SO(2)_D$ restores, one can always undo the twist.  This reverse procedure gives  
 the so-called emergent flavor  $SU(2) $  subgroup of $SU(2n_f)$  for free. 
To summarize,  the  compact lattice QED$_3$  possesses
\begin{equation}
{\cal G}_{{\rm QED}_3  } \sim 
  G_{\rm discrete} \times  C \times P \times T \times U(1)_V  \times  SU(n_f)_D 
\label{match1}
\end{equation}
This is indeed the symmetry structure of spin system in the gauge theory formulation in the $\pi$-flux state.  This needs to be compared with  much larger microscopic symmetry  (\ref{master1}) of P(F) theory. 
 
The  analog of the 
 $U(1)_A$ symmetry in the  P(F) theory is part of  the $SU(2) \subset SU(2n_f)$  symmetry  in 
 the continuum of the QED$_3$. 
 Unfortunately,   in the $\pi$-flux state of the 
   spin system,
and    in the specific lattice regularization  described above, 
   the continuous  $U(1)_A$ does not survive  at the cut-off  scale. Only  a 
   discrete subgroup of it is hidden in   $G_{\rm discrete}$. However,  $G_{\rm discrete}$  is practically useless (like any other discrete symmetry) for forbidding generic flux operators in lattice  QED$_3$.
  
 This is the  significant difference between P(F) theory, QCD(F)* theory and lattice QED$_3$.   The P(F) theory has $U(1)_A$ symmetry at short distances and this transmutes into a continuous topological symmetry in the IR preventing a mass term for a  photon, for any number of flavors.   In QCD(F)*, the short distance theory only has a $\Z_{2}$  discrete chiral symmetry,  
 which again transmutes into a trivial  
 $(\Z_{1})_{*}$  topological symmetry,  which  cannot prohibit mass term for the dual photon. For small numbers of flavors, the theory exhibits a mass gap in gauge sector.  At sufficiently large number of flavors, an accidental $U(1)_{*}$ may arise  as discussed in ($\S$.\ref{accidental}). 
 In the  lattice versus continuum QED$_3$, the critical target theory  has a $U(1)_A$ symmetry embedded into $SU(2n_f)$ for {\it any} $n_f$. However, the lattice  Hamiltonian  does not respect it.  This makes this problem different and relatively harder than  the previous two problems that we have discussed. 
 
\subsection{The emergent topological $U(1)_{*}$  symmetry} 
It is not { \it a priori } clear whether there is a  relation  between  P(F) theory, and lattice QED$_3$ studied  in   Ref.\cite{hermele-2004-70}. 
 Clearly, continuum P(F) theory is a theory with a scalar and with a  larger  set of symmetries than the lattice QED$_3$.  
However, the infrared physics of these  two theories seems to be coincident at least in  the large $n_f$ limit.  It is in principle plausible  that  different microscopic theories may flow to the same theory in their long distance limits. 

In our opinion, the most important physical issue is associated with the  topological 
$U(1)_{*}$ symmetry.
   In P(F), the origin of $U(1)_{*}$ is clear.  It is a natural consequence of the   the $U(1)_A$   symmetry combined with the  Callias index theorem. 
In large $n_f$ lattice QED$_3$,  the $U(1)_{*}$ symmetry is referred as an emergent topological symmetry 
of the IR theory \cite{hermele-2004-70}.  The reason it may be considered emergent is 
twofold: One is the analog of  $U(1)_A$ is not present in the spin system and resulting 
lattice QED$_3$. The second is the analog of the Callias index theorem on lattice QED$_3$ does not exist as shown by  Marston \cite{Marston:1990bj}.

The result of  Ref.~\cite{Marston:1990bj}  looks  discouraging,  as 
stated in \cite{kim-1999-272}. However, the more severe  issue is  the 
 absence of the $U(1)_A$ symmetry  in lattice QED$_3$, or spin system.  
 Below we will prove the following assertion: 
 If  the $U(1)_A$   is a symmetry of the cut-off (lattice) QED$_3$  theory,  despite the absence of the Callias index theorem, the topological $U(1)_{*}$ symmetry will emerge  in the long distances
 even at {\it small} $n_f$.

Let us see how this works. The result of 
ref.~\cite{Marston:1990bj}   does not tell us that   monopole-multifermion type operators are   excluded.  It only states that in a monopole  operator 
of the form $e^{i \sigma} O_{\rm fermion}$, the structure of $O_{\rm fermion}$ is not dictated 
by an index theorem. $O_{\rm fermion}$ may be 
$\{1, \;  (2 \; {\rm fermions}),\;  (4 \; {\rm fermions}),  \ldots \}$, a plethora of  (even) numbers of  fermion insertions allowed by other symmetries of the lattice. 
  Let us  list  a set of operators which may be induced nonperturbatively 
  \begin{equation}
\{ e^{i \sigma},  e^{i \sigma} \lambda_{1,a} \bar \lambda_{2,a},  \;\;  e^{i \sigma} (\lambda_{1,a} \bar \lambda_{2,a} )^2, \ldots,    e^{2 i  \sigma}, \;  e^{2 i  \sigma} \lambda_{1,a} \bar \lambda_{2,a}, \; 
 e^{2 i \sigma}  (\lambda_{1,a} \bar \lambda_{2,a})^2, \ldots  \}
 \label{fluxop}
 \end{equation}
 where we  suppress Lorentz  indices.
 This is the set of monopole operators  and the composites of monopoles with the fermion fields. By assumption, the $U(1)_A$, under which  
 $\lambda_{1,a} \bar \lambda_{2,a} \rightarrow e^{2i \beta} \;  \lambda_{1,a} \bar \lambda_{2,a} $
  is a symmetry of the cut-off theory. 
Our goal here is to  show that  the absence of an index theorem by itself does not imply that the continuous $U(1)_A$  symmetry cannot be  transmuted into the dual photon as a 
shift symmetry. 

  Let   $e^{i \sigma}   (\lambda_{1,a} \bar \lambda_{2,a})^{q} $ be the lowest dimensional flux operator   with multiple fermion insertions   allowed by lattice symmetries.  
  Since the $U(1)_A$ is a symmetry of the cut-off theory, it must be a symmetry of the long distance theory.  As before, this can be accomplished by intertwining $U(1)_A$ with 
  $U(1)_{\rm flux}$, the shift symmetry of dual photon,   in the infrared. 
 The invariance of  $e^{i \sigma}  (\lambda_{1,a} \bar \lambda_{2,a})^q$ under $U(1)_A$ demands  
that  the dual photon  must have a shift symmetry $\sigma  \rightarrow \sigma- 2 q \beta$. 
 Thus, reconciling $U(1)_A$ symmetry with the long distance physics  forbids any operators in the list except   $[e^{i \sigma}  (\lambda_{1,a} \bar \lambda_{2,a})^q]^k$. 
Most importantly, it forbids the monopole operator 
$e^{i \sigma} $ and other pure flux operators such as $e^{2 i \sigma} $
regardless of the value of $q \geq 1$.   
 This implies that   relevant monopole operators (which render the photon massive) may  be forbidden by the {\it accidental}      $U(1)_{*}$ pure flux forbidding 
 symmetry even at small $n_f$ if the cut-off theory has the $U(1)_A$ symmetry. 
  
Unfortunately,  the spin system does  not have the analog of $U(1)_A$ symmetry.  
The flux operators such as $e^{i \sigma}$ which are not forbidden by symmetry  will be generated.  Under the given circumstances, the {\it only} way that such operators will not  generate a  mass  for dual photon is 
if they are irrelevant  in the long distances in the renormalization group sense.  We reach to the 
conclusion that, for spin systems in a $\pi$-flux phase,  unlike the P(F) theories, the renormalization group and large $n_f$ analysis  are  unavoidable  \cite{hermele-2004-70}.

\section{Conclusions and prospects}
\begin{table}[htdp]
\begin{center}
\begin{tabular}{| p{2.5cm} |p{4cm}|p{2cm}|c|p{2cm}|}
\hline
  Theory  & Description & Topological  symmetry,   microscopic precursor  & Gauge sector  & Long \qquad distances \\
\hline 
 P \cite{Polyakov:1976fu}
 & noncompact $\Phi $,  & none, none  & gapped & confined  \\ \hline 
P(adj) \cite{Affleck:1982as}  & noncompact $\Phi$, real ${\cal R}$,  complex fermions & $U(1)_{*}, U(1)_A$ & gapless & deconfined, free photon  \\ 
\hline  P(F) &noncompact $\Phi$, complex ${\cal R}$, complex fermions &  $U(1)_{*}, U(1)_A$ & gapless &  deconfined, CFT \\ \hline
YM*   \cite{Unsal:2008ch}& compact  $\Phi$ & none, none  & gapped & confined  \\ \hline 
QCD(adj)* \cite{Unsal:2007jx}  & compact $\Phi$, real ${\cal R}$, complex fermions, 
$n_f$ small  & $(\Z_{N})_{*}$, $\Z_{2N}$ axial & gapped & confined   \\ 
\hline  QCD(F)* \cite{Shifman:2008ja} & compact $\Phi$, complex ${\cal R}$, complex fermions, $n_f$ small 
& none, $\Z_2$ & gapped &  confined 
 \\ \hline
compact  lattice QED$_3$ \cite{Polyakov:1976fu}
 & 
 compact gauge fluctuations
 & none,  none   & gapped  &  confined
 \\ 
 \hline
compact  lattice QED$_3$ with fermions 
 \cite{hermele-2004-70}
 & 
 complex ${\cal R}$, complex fermions,  $N_f \gg 1$ & emergent $U(1)_{*}$, 
 none   & gapless  &  deconfined, CFT \\
 \hline
\end{tabular}
\end{center}
\caption{The role of topological symmetry in the determination of the deconfined/confined long distance behavior.  It is worth emphasizing that all the theories in the list has magnetic monopoles in a semi-classically tractable regime. Thus, the presence or absence of the 
magnetic monopoles does not tell much about the infrared property of the theory.  
   A more refined characterization is through the topological symmetry.}
\label{default}
\end{table}%
 
{\bf Topological symmetry and  classification of gauge theories:} 
In this paper, we discussed a large   class of gauge theories  formulated on  $\R^3$ and 
 $S^1 \times \R^3$ whose   long distance gauge structure is described by  abelian $U(1)^{N-1}$. 
Examples are  $SU(N)$ continuum  P(${\cal R}$ )  on $\R^3$, $SU(N)$ continuum 
QCD(${\cal R}$)*,  and  $U(1)^{N-1}$ lattice  QED$_3$ in three dimensions. We arrived to  sharp 
 topological symmetry realizations 
 which distinguish  the zero temperature phases of such gauge theories, such as confined versus 
 deconfined.  \footnote{In  
  QCD(${\cal R}$)*, the small $S^1 \times \R^3$ should be viewed as a spatial (not thermal) compactification, along which fermions are endowed with periodic boundary condition.  Its   Minkowski space continuation is $S^1 \times \R^{2,1}$.  If one wishes to study these gauge theories at finite temperature, a thermal circle should be formed out of  the temporal direction on $\R^{2,1}$.}   
 \begin{itemize} 
{\item[$\bf 1)$]The existence  of continuous $U(1)_{*}$ topological  symmetry is the 
necessary and sufficient condition to demonstrate  the  absence of mass gap  in the gauge sector and provides  an unambiguous characterization of de-confinement. }
\begin{itemize}
{\item[$\bf 1.a)$] If the $U(1)_{*}$ symmetry is spontaneously broken, then there is a Goldstone boson. The infrared theory is the free scalar (which is same as a photon on 
$\R^3$.)}
{\item[$\bf 1.b)$] If the $U(1)_{*}$ symmetry is  unbroken, the unbroken $U(1)_{*}$ protects the masslessness of the dual scalar.  In some cases, the infrared theory flows into an interacting  CFT. }
\end{itemize}
{\item[$\bf 2)$] 
The existence of a  discrete topological symmetry is necessary, but not sufficient
to exhibit confinement.}
\begin{itemize}
 {\item[$\bf 2.a)$] 
If the monopole (or other flux) operators are  irrelevant at large distances, then there is an emergent topological  $U(1)_{\rm flux}$   symmetry. This class of theories will deconfine, and some will 
flow into interacting CFTs.}  
 {\item[$\bf 2.b)$] 
If the monopole (or other flux) operator is  relevant at large distances, then the mass gap and confinement will occur. Showing the relevance of flux operators is the sufficient criteria to exhibit mass gap and confinement. 
}
\end{itemize}
\end{itemize}
Some examples for these classes are tabulated in table.\ref{default} along with useful references.
I wish to point out  that some of  these necessary and sufficient conditions are not 
completely novel. 
An example of class $1.a)$ was discussed long ago by Affleck, Harvey and Witten 
\cite{Affleck:1982as}, 
and the statement of $2.a)$ is constructed in  the work of 
Hermele et.al  \cite{hermele-2004-70} on  stable spin liquids, but it applies more generally to
 gauge theories.  The totality of these criteria is new. 
 \footnote{See also  Refs.~\cite{Hands:2006dh,  Di Giacomo:1999fa} 
 which use disorder operators to probe confinement. These works also attempt to provide a symmetry realization for confinement.  An application to a QCD-like theory 
 with adjoint fermions is  given in \cite{Cossu:2008wh}. It may be useful 
 to perform the lattice simulations on an  asymmetric lattice,
  which mimics $\R^3 \times S^1$ where $S^1$ is endowed with periodic spin connection for fermions.  The theoretical analysis shows that the small $S^1$   regime must exhibit confinement without chiral symmetry breaking \cite{Unsal:2007vu, Unsal:2007jx}. It would be interesting to test this on lattice.}

There are many interesting questions on  the generalizations 
of these  criteria.  The most obvious is  
whether the topological symmetry characterization  can be generalized to cases where the 
long distance dynamics is non-abelian.   
Another one  is whether the  abelian CFTs
discussed in this paper has non-abelian counterparts? Assuming this is the case, 
are they dual  to non-abelian spin liquids at large distances?  Can we make use of this topological characterization towards the decompactification $\R^4$ limit of QCD(${\cal R}$)*?
We leave these questions for future work.

{\bf Ambiguity in defining compact QED$_3$ in continuum and resolution:} 
  There are at least two continuum gauge theories which produce compact QED$_3$ in perturbation theory via gauge symmetry breaking in  P$({\cal R})$  and QCD$({\cal R})$*.
 These flow into opposite  IR  theories, such as a CFT versus a theory with a mass gap in 
 some cases,  as shown in table.\ref{default}.

{\bf Spin liquid and P(F) duality:} We demonstrated  that the $SU(N)$ 
 Polyakov model  with $2n_f$ massless fundamental fermions and $SU(n_f)_D$ spin systems in the  $n_f\gg 1 $ limit   flow into the  same  interacting conformal field theory. 
  This is to some extent surprising   due to the absence of the Callias index theorem in lattice QED$_3$ \cite{Marston:1990bj}, and very distinct symmetries of the spin Hamiltonian and P(F) model.
   Both theories are  quantum critical in the sense that 
 there are no  relevant perturbative or non-perturbative  operators consistent with  the symmetries of the microscopic theory.  Thus, these  theories flow into interacting conformal  field theories at long distances.  As the number of flavors is reduced, the long distance limit of $2n_f \geq 4$ 
P(F) theory   interpolate 
in between the weakly  and  strongly coupled  CFT's. What happens  with lattice QED$_3$ at small number of flavors is still ambiguous. 

Given the long distance duality between the spin liquids and P(F) gauge theory, 
a sensible question is the meaning of the doping of spin liquids by holons on the gauge theory side. Clearly, compactification of the field space brings in new excitations (flux operators) from infinity, and generates a QCD* type of theory, with a mass gap in its gauge sector. It is desirable to understand the relation, if any, between the QCD* theories and $d$-wave superconducting 
phase of high  $T_c$ cuprates.

\acknowledgments
I am grateful to  Eun-Ah Kim, B. Marston, M. Shifman  for multiple useful explanations. 
I also would like to thank  M. Headrick, S. Kachru,  M. Mulligan, \"O. Oktel,  T. Senthil,  
Piljyin Yi for related discussions.
This work was supported by the U.S.\ Department of Energy Grants DE-AC02-76SF00515.

\bibliographystyle{JHEP} 

\bibliography{confinement1}

\end{document}